\global\def\extension{1}      

\catcode`\@=11

\newif\if@fewtab\@fewtabtrue
\def\ps@draft{\let\@mkboth\@gobbletwo
    \def\@oddhead{}
    \def\@oddfoot{\hbox to 7 cm{\tiny \versionno
       \hfil}\hskip -7cm\hfil\rm\thepage \hfil}
    \def\@evenhead{}\let\@evenfoot\@oddfoot}

\def\draftcite#1{\ifnum\draftcontrol=1#1\else{}\fi}

\def\@lbibitem[#1]#2{\item{}\hskip -3cm \hbox to 2cm
{\hfil$\scriptstyle\draftcite{#2}$}\hskip
1cm[\@biblabel{#1}]\if@filesw
     {\def\protect##1{\string ##1\space}\immediate
      \write\@auxout{\string\bibcite{#2}{#1}}}\fi\ignorespaces}

\def\@bibitem#1{\item\hskip -3cm \hbox to 2cm
{\hfil {\footnotesize\draftcite{#1}}}\hskip 1cm
\if@filesw \immediate\write\@auxout
       {\string\bibcite{#1}{\the\value{\@listctr}}}\fi\ignorespaces}

\ifnum\extension=0 \newcommand\internal[1]{} \fi
\ifnum\extension=1 \newcommand\internal[1]{\smallskip
\noindent\rule{.4em}{.4em} \hsp{-1} \raisebox{.28em}{\rule{14em}{.1em}}
{\footnotesize\sf Technical details:}
\raisebox{.28em}{\rule{14em}{.1em}} \hsp{-1} \rule{.4em}{.4em}
\\[.3em] {\small #1}\\[.3em]
\rule{.4em}{.4em} \hsp{-1} \rule{12.8em}{.1em}
{\footnotesize\sf End of technical details.}
\rule{12.8em}{.1em} \hsp{-1} \rule{.4em}{.4em}
\smallskip} \fi
  \def\versionno{ SO2 -- version 10.5 -- by jf --     21.2.96  }

\def\citen#1{\if@filesw \immediate\write \@auxout {\string\citation{#1}}\fi%
\@tempcntb\m@ne \let\@h@ld\relax \def\@citea{}%
\@for \@citeb:=#1\do {\@ifundefined {b@\@citeb}%
    {\@h@ld\@citea\@tempcntb\m@ne{\bf ?}%
    \@warning {Citation `\@citeb ' on page \thepage \space undefined}}%
    {\@tempcnta\@tempcntb \advance\@tempcnta\@ne
    \setbox\z@\hbox\bgroup\ifcat0\csname b@\@citeb \endcsname \relax
    \egroup \@tempcntb\number\csname b@\@citeb \endcsname \relax
    \else \egroup \@tempcntb\m@ne \fi \ifnum\@tempcnta=\@tempcntb
    \ifx\@h@ld\relax \edef \@h@ld{\@citea\csname b@\@citeb\endcsname}%
    \else \edef\@h@ld{\hbox{--}\penalty\@highpenalty
    \csname b@\@citeb\endcsname}\fi
    \else \@h@ld\@citea\csname b@\@citeb \endcsname \let\@h@ld\relax \fi}%
\def\@citea{,\penalty\@highpenalty\hskip.13em plus.13em minus.13em}}\@h@ld}
\def\@citex[#1]#2{\@cite{\citen{#2}}{#1}}%
\def\@cite#1#2{\leavevmode\unskip\ifnum\lastpenalty=\z@\penalty\@highpenalty\fi%
  \ [{\multiply\@highpenalty 3 #1%
  \if@tempswa,\penalty\@highpenalty\ #2\fi}]}   %
\makeatother 

\def\aff           {affine Lie algebra}
\def\aft           {algebraic field theory}
\def\alg           {algebra}

\def\AW            {\mbox{${\mathfrk A}_{\scriptstyle\rm WZW}$}}
\renewcommand\b[3] {b^{#1;#2}_{#3}}
\newcommand\BX[4]  {B^{#1,#2;#3}_{#4}}

\def\be            {\begin{equation}}
\def\Be            {$}
\def\bearl         {\begin{array}{l}}
\def\bearll        {\begin{array}{ll}}
\def\bearlll       {\begin{array}{lll}}
\def\bfe           {{\bf1}}

\def\car           {canonical an\-ti-com\-mu\-ta\-tion relation}
\def\cara          {CAR algebra}

\newcommand\cbm[2] {\bar c^{#1,-}_{#2}}
\newcommand\cbp[2] {\bar c^{#1,+}_{#2}}
\newcommand\cbpm[2]{\bar c^{#1,\pm}_{#2}}

\def\cft           {conformal field theory}
\def\cfts          {conformal field theories}

\def\CKG           {{\cal C}({\cal K},\Gamma)}
\def\CKGh          {{\cal C}(\hat{\cal K},\hat{\Gamma})}

\newcommand\ceps[2]{c^{#1,\eps}_{#2}}
\newcommand\ceta[2]{c^{#1,\eta}_{#2}}
\def\chiCc         {\raisebox{.15em}{$\chi$}\Coset_\rmc} 
 
\def\chicjl        {\raisebox{.15em}{$\chi$}^{{\rm c};\scriptstyle \el}_{[j]}} 
\newcommand\chicjzN[1]{\raisebox{.15em}{$\chi$}^{{\rm c};
                   \scriptstyle 2N}_{[#1]}} 
\def\chiCj         {\raisebox{.15em}{$\chi$}\Coset_{[j]}} 
\def\chiCm         {\raisebox{.15em}{$\chi$}\Coset_{[M]}} 
\def\chicol        {\raisebox{.15em}{$\chi$}^{{\rm c};\scriptstyle \el}\Null} 
\def\chicozN       {\raisebox{.15em}{$\chi$}^{{\rm c};\scriptstyle 2N}\Null} 
\def\chiCo         {\raisebox{.15em}{$\chi$}\Coset\Null} 
\def\chiCoM        {\raisebox{.15em}{$\chi$}^{{\rm c};\scriptstyle M}\Null} 
\def\chiCs         {\raisebox{.15em}{$\chi$}\Coset_\rms} 
 
\def\chicvl        {\raisebox{.15em}{$\chi$}^{{\rm c};\scriptstyle \el}_\rmv} 
\def\chicvzN       {\raisebox{.15em}{$\chi$}^{{\rm c};\scriptstyle 2N}_\rmv}  
\def\chiCv         {\raisebox{.15em}{$\chi$}\Coset_\rmv}   
\def\chieo         {\raisebox{.15em}{$\chi$}\ke\Null} 
\def\chiev         {\raisebox{.15em}{$\chi$}\ke_\rmv}  

\def\chiJ          {\raisebox{.15em}{$\chi$}\ns_J}
\newcommand\chivir[1]{\raisebox{.15em}{$\chi$}^{\rm Vir}_{#1}}
\newcommand\chim[1]{\raisebox{.15em}{$\chi$}\ns_{[#1]}}
\def\chio          {\raisebox{.15em}{$\chi$}\ns_0}
\def\chizc         {\raisebox{.15em}{$\chi$}\kz_\rmc} 
\def\chizj         {\raisebox{.15em}{$\chi$}\kz_{[j]}} 
\def\chizl         {\raisebox{.15em}{$\chi$}\kz_{[\el]}} 
\def\chizo         {\raisebox{.15em}{$\chi$}\kz\Null} 
\def\chizs         {\raisebox{.15em}{$\chi$}\kz_\rms} 
\def\chizv         {\raisebox{.15em}{$\chi$}\kz_\rmv}

\def\CKGIh         {{\cal C}(\hat{\cal K}(I),\hat{\Gamma})}
\newcommand\cm[2]  {c^{#1,-}_{#2}}

\def\cocon         {coset construction}
\def\complex       {{\dl C}}
\def\coset         {^{\rm c}}
\def\Coset         {^{{\rm c};\scriptscriptstyle M}}

\def\cost          {\cos(t)\,}

\newcommand\cp[2]  {c^{#1,+}_{#2}}
\newcommand\cpm[2] {c^{#1,\pm}_{#2}}
\def\csa           {Cartan subalgebra}
\def\cwb           {Cartan\hy Weyl basis}

\newcommand\del[2] {\delta_{#1,#2}}
\newcommand\Delns[1]{\Delta\Ns_{n;#1}}
\newcommand\Delnsb[1]{\bar \Delta\Ns_{n;#1}}
\newcommand\Delc[1]{\Delta^{\!\rm c}_{n;#1}}
\newcommand\Delcb[1]{\bar \Delta^{\!\rm c}_{n;#1}}
\def\dh            {\Delta}
\def\Dn            {\mbox{${\cal D}_N$}}

\let\dstyle=\displaystyle
\def\dsum          {\displaystyle\sum}

\def\ee            {\end{equation}}
\def\eE            {{\rm e}}
\def\Ee            {$ }
\newcommand\EE[2]  {{\cal E}^{#1}_{#2}}
\def\eear          {\end{array}}
\def\eit           {{\rm e}^{{\rm i}t}}
\def\el            {\ell}

\def\emit          {{\rm e}^{-{\rm i}t}}
\def\emt           {energy-momentum tensor}
\newcommand\EO[2]  {E^{#1}_{#2}}
\let\eps=\varepsilon
\newcommand\erf[1] {(\ref{#1})}
\def\fA            {\mathfrk{A}}
\def\fAg           {\mathfrk{A}}
\def\fF            {\mathfrk{F}}
\def\fFg           {\mathfrk{F}}
\def\findim        {finite-dimensional}
\def\fline         {{~}\\[1 mm]\noindent ------------------\\[1 mm]}

\def\forzl         {{\rm for}\ N=2\el}
\def\forzle        {{\rm for}\ N=2\el+1}
\newcommand\Frac[2]{\mbox{\large$\frac{#1}{#2}$}}
\def\fsqz          {\mbox{\large$\frac1{\sqrt2}$}\,}

\def\fstar         {\star}
\def\futnot#1      {\ifnum\draftcontrol=1%
                   \footnote{~{\sc internal footnote:} #1}\ \fi}
\def\futnote#1     {\footnote{~#1}\ }
\def\g             {\mbox{$\mathfrk g$}}
\def\gb            {\mbox{$\bar{\mathfrk g}$}}
\def\GG            {\Gamma}
\def\GGh           {\hat{\Gamma}}

\def\h             {{1/2}}
\def\half          {\mbox{\large$\frac12$}\,}
\def\halfi         {\mbox{\large$\frac\ii2$}\,}
\newcommand\Hh[1]  {{\cal H}_{#1}}
\newcommand\HH[1]  {{\cal H}^{#1}_{}}
\def\Hheo          {{\cal H}_\circ\ke}
\def\Hhev          {{\cal H}_\rmv\ke}
\def\Hhkvl         {{\cal H}_{\Lambda}\kkv}

\def\Hhzc          {{\cal H}_\rmc\kz}
\def\Hhzj          {{\cal H}_{[j]}\kz}
\def\Hhzl          {{\cal H}_{\Lambda}\kz}
\def\Hhzo          {{\cal H}_\circ\kz}
\def\Hhzs          {{\cal H}_\rms\kz}
\def\Hhzv          {{\cal H}_\rmv\kz}
\newcommand\Hhh[1] {{\cal H}_{[#1]}}
\def\Hhj           {{\cal H}_J}

\def\Hhm           {{\cal H}_{[m]}}
\def\Hho           {{\cal H}_0}
\def\HNS           {{\cal H}_{\rm NS}}
\def\HNSh          {\hat{\cal H}_{\rm NS}}
\def\hsa           {horizontal subalgebra}
\newcommand\hsp[1] {\mbox{\hspace{#1em}}}
\def\hw            {highest weight}
\def\hwm           {highest weight module}

\def\hws           {highest weight state}
\def\hwv           {highest weight vector}
\def\hy            {$\mbox{-\hspace{-.66 mm}-}$}
\def\id            {\mbox{\sl id}}
\def\ide           {identification}

\def\ii            {{\rm i}}
\def\ihwm          {irreducible highest weight module}
\def\iN            {\!\in\!}

\def\Infdim        {Infinite dimensional}
\def\Intt          {\Frac1{2\pi}\dstyle\int_0^{2\pi}\!\!{\rm d}t\,}
\def\IntT          {\Frac1{4\pi}\dstyle\int_0^{2\pi}\!\!{\rm d}t\,}
\def\intt          {\Frac1{2\pi}\int_0^{2\pi}\!\!{\rm d}t\,}
\def\intT          {\Frac1{4\pi}\int_0^{2\pi}\!\!{\rm d}t\,}
\def\irmod         {irreducible module}
\def\irrep         {irreducible representation}
\newcommand\J[3]   {J^{#1,#2}_{#3}}
\newcommand\Jm[2]  {J_{m}^{#1,#2}}
\newcommand\Jmt[4] {J_{m}(\te{#1}{#2}{#3}{#4})}
\newcommand\Jmtee[2]{J_{m}(\ttee{#1}{#2})}
\newcommand\Jmte[1]{J_m(t^{#1}_{\eps})}
\newcommand\Jmttm[1]{J_m(t^{#1}_-)}
\newcommand\Jmttp[1]{J_m(t^{#1}_+)}
\newcommand\Jmttpm[1]{J_m(t^{#1}_{\pm})}
\newcommand\Jo[2]  {J_{0}^{#1,#2}}
\newcommand\Jot[4] {J_{0}(\te{#1}{#2}{#3}{#4})}
\newcommand\Jt[5]  {J_{#1}(\te{#2}{#3}{#4}{#5})}

\def\Jzeta         {{\cal J}_\zeta}
\def\ke            {^{\scriptscriptstyle(1)}}
\def\KK            {{\cal K}}
\def\KKh           {\hat{\cal K}}
\def\kkv           {^{\scriptscriptstyle(k^\vee)}}
\def\kma           {Kac\hy Moo\-dy algebra}

\def\kv            {\mbox{$k_{}^{\scriptscriptstyle\vee}$}}
\def\kV            {{k_{}^{\scriptscriptstyle\vee}}}
\def\kz            {^{\scriptscriptstyle(2)}}
\long\def\labl#1   {\label{#1}\ee \ifnum\draftcontrol=1 
                   \mbox{ }\\[-12 mm]\query{#1}\\[5 mm] \fi}
\def\lc            {\Lambda_{{\rm c}}}

\def\lie           {Lie algebra}

\newcommand\lj[1]  {\Lambda_{[#1]}}
\newcommand\Lj[1]  {\Lambda_{(#1)}}
\def\llb           {\mbox{\large[}}
\def\lLb           {\mbox{\large(}}
\def\Llb           {\mbox{\Large\{}}
\def\LLb           {\mbox{\Large[}}
\newcommand\LNS[1] {\mbox{$L_{#1}\ns$}}
\def\lo            {\Lambda\Null}
\def\lrb           {\mbox{\large]}}
\def\lRb           {\mbox{\large)}}
\def\LRb           {\mbox{\Large]}}
\def\Lrb           {\mbox{\Large\}}}
\def\ls            {\Lambda_{{\rm s}}}
\def\lv            {\Lambda_{{\rm v}}}
 
\def\mathbb        {\dl }
\def\mh            {{-1/2}}
\def\mnh           {{-n-1/2}}

\def\natnum        {{\dl N}}
\def\natnumo       {{\dl N}_0}

\def\nh            {{n+1/2}}
\newcommand\nline[1] {\\{}\\[-.#1em]}
\def\nmh           {{n-1/2}}

\newcommand\normord[1] {\,{\bf:}#1{\bf:}\,}
\def\ns            {^{\scriptscriptstyle\rm NS}}
\def\Ns            {^{\!\scriptscriptstyle\rm NS}}

\def\NS            {Neveu\hy Schwarz }
\def\Null          {_\circ}

\newcommand\Ocnpm[1] {|\Omega^{#1,\pm}_{\rm c}\rangle}
\newcommand\Ocnmp[1] {|\Omega^{#1,\mp}_{\rm c}\rangle}

\def\Oe            {{\rm O}(1)}
\def\Ok            {{\rm O}(\kv)}
\newcommand\om[1]  {|\Omega^{}_{#1}\rangle}

\def\OM            {\om{}}

\newcommand\Ombnmp[2]{|\overline\Omega^{#2,\mp}_{[#1]}\rangle}

\newcommand\Ombnpm[2]{|\overline\Omega^{#2,\pm}_{[#1]}\rangle}

\newcommand\OmfJo[1]{|\Omega^{J,#1}\Null\rangle}
\newcommand\OmfJv[1]{|\Omega^{J,#1}_\rmv\rangle}
\newcommand\OmfOo[1]{|\Omega^{0,#1}\Null\rangle}
\newcommand\OmfOv[1]{|\Omega^{0,#1}_\rmv\rangle}

\newcommand\Omp[1] {|\Omega^+_{[#1]}\rangle}

\newcommand\Omnmp[2]{|\Omega^{#2,\mp}_{[#1]}\rangle}

\newcommand\Omnpm[2]{|\Omega^{#2,\pm}_{[#1]}\rangle}

\def\one           {\mbox{\small $1\!\!$}1}
\def\onedim        {one-dimensional}
\def\onehalf       {\mbox{$\frac12$}}
\def\onetol        {1,2,...\,,\el}
\def\onetolme      {1,2,...\,,\el-1}
\def\onetolmz      {1,2,...\,,\el-2}
\def\onetom        {1,2,...\,,M}
\def\onetomme      {1,2,...\,,M-1}
\def\oneton        {1,2,...\,,n}
\def\onetoN        {1,2,...\,,N}

\def\ONSh          {|\hat\Omega_{\rm NS}\rangle}
\newcommand\Oonmp[1] {|\Omega^{#1,\mp}\Null\rangle}
\newcommand\Oonpm[1] {|\Omega^{#1,\pm}\Null\rangle}
\newcommand\Osnpm[1] {|\Omega^{#1,\pm}_{\rm s}\rangle}
\newcommand\Osnmp[1] {|\Omega^{#1,\mp}_{\rm s}\rangle}

\def\otol          {0,1,...\,,\el}
\def\otolme        {0,1,...\,,\el-1}
\def\Ovvo           {|\Omega_{{\rm v}}\rangle}
\def\Ovo           {|\Omega^{0,\pm}_{{\rm v}}\rangle}

\newcommand\Ovnpm[1] {|\Omega^{#1,\pm}_{\rm v}\rangle}
\newcommand\Ovnmp[1] {|\Omega^{#1,\mp}_{\rm v}\rangle}
\def\Oz            {{\rm O}(2)}
\def\pfc           {\phi_\rmc}
\newcommand\pfj[1] {\phi_{[#1]}}
\def\pfo           {\phi\Null}
\def\pfs           {\phi_\rms}
\def\pfv           {\phi_\rmv}
\def\Phila         {|\Phi_\Lambda\rangle}

\def\PiNSh         {\hat\pi_{\rm NS}}
\def\PNS           {P_{\rm NS}}
\def\PNSh          {\hat{P}_{\rm NS}}

\def\PJ            {\mbox{$P^{}_{\!J}$}}
\def\Po            {\mbox{$P^{}_{\!0}$}}
\newcommand\Pm[1]  {\mbox{$P^{-}_{\![#1]}$}}
\newcommand\Pp[1]  {\mbox{$P^{+}_{\![#1]}$}}
\newcommand\Ppm[1] {\mbox{$P^{\pm}_{\![#1]}$}}

\def\psiemq        {\mbox{$\psi^{}_{1,0}(-q)$}}
\newcommand\psim[1]{\mbox{$\psi^{}_{M,#1}$}}

\newcommand\psinh[1]{\mbox{$\psi^{}_{\el,#1}$}}
\newcommand\psinz[1]{\mbox{$\psi^{}_{2N,#1}$}}

\def\qdi           {\mbox{$\cal D$}}

\def\qfts          {quantum field theories}
\def\qk            {1,2,...\,,\kv}
\def\QKG           {{\cal Q}({\cal K},\Gamma)}
\def\qlo           {q^{L_0^{(\rm NS)_{}}}}
\def\qmh           {q^{m+1/2}}
\long\def\query#1{\hskip 0pt{\vadjust{\everypar={}\small\vtop to 0pt{\hbox{}%
     \vskip -13pt\rlap{\hbox to 47pc{\hfil{\vtop{\hsize=8pc\tolerance=6000%
     \hfuzz=.5pc\rightskip=0pt plus 3em\noindent#1}}}}\vss}}}}%
\def\qzme          {q^{2m+1}}

\def\reals         {{\dl R}}
\newcommand\Reh[1] {W(1,#1)}
\def\rep           {representation}
\def\Rep           {Representation}
\def\resp          {respectively}

\def\rmc           {{\rm c}}
\def\rms           {{\rm s}}
\def\rmo           {\circ}
\def\rmv           {{\rm v}}
\def\Rn            {\mbox{${\cal R}\kz_{\rm NS}$}}
\def\Ro            {\mbox{${\cal R}^{}_{\Oz}$}}
\def\role          {r\^ole}
\newcommand\rp[1]  {\Phi_{[#1]}}
\def\rpj           {\Phi_J}
\def\rpo           {\Phi_0}
\def\Rw            {\mbox{${\cal R}\kz_{\rm WZW}$}}
\newcommand\sect[1] {\section{#1}\setcounter{equation}{0}}
\newcommand\Sect[2] {\sect{#1}\label{s.#2}
                   \ifnum\draftcontrol=1 \query{s.#2} \fi}
\def\seta          {\eta}
\def\sgam          {\gamma}
\def\sgamt         {\gamma_t}

\def\sint          {\sin(t)\,}

\newcommand\smallmatrix[1] {\mbox{\footnotesize $\left(\begin{array}#1
                   \end{array}\right)$}}
\newcommand\sN[1]  {\Theta^{}_{N,#1}(q)}
\def\son           {\mbox{so$(N)$}}
\def\sonh          {\mbox{$\widehat{\rm so}(N)$}}
\def\sonhe         {\mbox{$\widehat{\rm so}(N)_1$}}
\def\sonhz         {\mbox{$\widehat{\rm so}(N)_2$}}

\newcommand\srf[1] {section \ref{s.#1}}

\def\subseT        {\!\subset\!}
\newcommand\summN[1]{{\displaystyle\sum_{\scriptstyle m_1,m_2,...,m_N\in\zet
                   \atop \scriptstyle m_1+m_2+...+m_N=#1}\!\!\!}}
\newcommand\sumMN[1]{{\displaystyle\sum_{\scriptstyle \vecm\in\zet^N
                   \atop \scriptstyle \sum m_i=#1}\!}}
\def\summZ         {\sum_{m\in\zet}}
\newcommand\sumni[1]{\sum_{n=#1}^\infty}
\def\sumnZ         {\sum_{n\in\zet}}
\def\sumq          {{\displaystyle\sum_{q=1}^2}}
\def\sumqv         {{\displaystyle\sum_{q=1}^\kV}}
\def\sun           {\mbox{su$(N)$}}
\def\sunh          {\mbox{$\widehat{\rm su}(N)$}}
\def\sunhe         {\mbox{$\widehat{\rm su}(N)_1$}}
\def\suse          {superselection sector}
\newcommand\T[2]   {T^{#1,#2}}
\newcommand\te[4]  {t^{#1,#2}_{#3,#4}}

\def\tp            {\times}

\newcommand\tte[1] {t^{#1}_{\eps}}
\newcommand\ttee[2]{t^{#1,#2}_{\eps,\eta}}

\newcommand\ttp[1] {t^{#1}_+}
\newcommand\ttpm[1]{t^{#1}_{\pm}}
\def\trNS          {{\rm tr}^{}_{\HNSh}\!}
\def\twodim        {two-di\-men\-si\-o\-nal}

\def\Uc            {\Hhzc}  
\def\Uget          {\mbox{$U(\seta\sgamt)$}}
\def\Ugt           {\mbox{$U(\sgamt)$}}
\def\Uj            {\Hhzj}  
\def\Uk            {{\rm U}(\kv)}
\def\unh           {\mbox{$\widehat{\rm u}(N)$}}
\def\Uo            {\Hhzo}  
\def\Us            {\Hhzs}  
\def\Uv            {\Hhzv}  

\def\vecm          {{\bf m}}

\def\vi            {\varphi}

\def\wi            {\xi}
\def\wrt           {with respect to }
\def\wrtt          {with respect to the }
\def\WZW           {Wess\hy Zumino\hy Witten}

\def\wzwt          {WZW theory}
\def\wzwts         {WZW theories}
\newcommand\x[2]   {x^{#1,+}_{#2}}
\newcommand\xpm[2] {x^{#1,\pm}_{#2}}
\newcommand\xmp[2] {x^{#1,\mp}_{#2}}

\newcommand\Xpm[2] {X^{#1,\pm}_{#2}}
\newcommand\Xmp[2] {X^{#1,\mp}_{#2}}
\newcommand\xb[2]  {\bar x^{#1,+}_{#2}}
\newcommand\xbpm[2]{\bar x^{#1,\pm}_{#2}}
\newcommand\xbmp[2]{\bar x^{#1,\mp}_{#2}}

\newcommand\Xbpm[2]{\bar X^{#1,\pm}_{#2}}
\newcommand\Xbmp[2]{\bar X^{#1,\mp}_{#2}}
\newcommand\y[2]   {x^{#1,-}_{#2}}

\newcommand\yb[2]  {\bar x^{#1,-}_{#2}}

\def\zet           {{\dl Z}}

\global\def\draftcontrol{0}
\catcode`\@=12

\documentclass[12pt]{article}
\usepackage{amssymb,amsfonts}

\begin{document}
\setlength{\unitlength}{.1 em}

\let\dl=\bf \let\mathfrk=\bf
\let\dl=\mathbb  \let\mathfrk=\mathfrak


\begin{flushright}  {~} \\[-15 mm]  {\sf hep-th/9602116} \\[1mm]
{\sf DESY 96-030} \\[1 mm]{\sf February 1996} \end{flushright} \vskip 2mm

\begin{center} \vskip 14mm
{\Large\bf HIGHER LEVEL WZW SECTORS}\\[3.2mm] {\Large\bf FROM FREE FERMIONS}\\
\vskip 15mm
{\large Jens B\"ockenhauer $^{\sf D}$} \\[5mm]
{II.\ Institut f\"ur Theoretische Physik, Universit\"at Hamburg\\[1mm]
Luruper Chaussee 149, \ D -- 22761~~Hamburg} \\[9mm]
{\large J\"urgen Fuchs $^{\sf H}$} \\[5mm]
{DESY\\[1mm] Notkestra\ss e 85, \ D -- 22603~~Hamburg} 
\end{center} \vskip 22mm

\begin{quote}{\bf Abstract}.\\
We introduce a gauge group of internal symmetries of an ambient algebra
as a new tool for investigating the superselection structure of WZW 
theories and the representation theory of the corresponding affine Lie 
algebras. The relevant ambient algebra arises from the description of 
these conformal field theories in terms of free fermions. As an 
illustration we analyze in detail the \son\ WZW theories at level two. 
In this case there is actually a homomorphism from the representation 
ring of the gauge group to the WZW fusion ring, even though the level-two 
observable algebra is smaller than the gauge invariant subalgebra of the 
field algebra.  \end{quote}
\vfill {}\fline{} {\small 
$^{\sf D}$~~Supported by Deutsche Forschungsgemeinschaft\\[.4em]   
$^{\sf H}$~~Supported by Deutsche Forschungsgemeinschaft (Heisenberg fellow)} 
\newpage


\sect{Introduction}

While a wealth of information about \wzwts\ has been obtained by 
analyzing these \cfts\ with the help of the unbounded operators which
generate their Virasoro and affine \lie\ structures,
much less is known about the superselection structure of \wzwts\ as
described in terms of local algebras of bounded operators.
In comparison with higher-dimensional relativistic \qfts, 
some difficulties arise in these models as a consequence of the fact that
the quantum symmetry which governs the
superselection structure is not a gauge group in the sense of
Doplicher, Haag and Roberts (DHR \cite{dohr}). So far no generally
accepted description of this quantum symmetry is available.
Accordingly, the analysis of WZW models in the framework of
\aft\ has been confined to the case of level one of the relevant affine \lie s
\cite{fugv,bock3} or to simple currents \cite{fugv2}, i.e.\ 
sectors with unit statistical dimension.%
\,\futnote{Similar remarks apply to the work on other \cfts, compare e.g.\
\cite{bumt,masc,rehr11,bock}. For an approach which addresses general \cfts,
see \cite{wass?}.} Here we report on ideas 
which allow to deal also with more complicated situations.

The purpose of this paper is twofold. First, we would like to find a 
convenient substitute for the DHR gauge group in
low-dimensional field theories. We do {\em not\/} require that this 
substitute plays the \role\ of the full quantum symmetry of the theory, i.e.\ 
the gauge invariant fields need not coincide with the observables of the
theory under consideration, so that the gauge group does not directly describe 
the superselection structure. Rather, we only
demand that it supplies a tool for examining this structure, which
when combined with other information allows to characterize the sectors at
least to a large extent. In the specific case of \wzwts, the required 
additional information comes from the \rep\ theory of \aff s.
In this case we succeed in identifying a symmetry group which
satisfies the required property.

Our second goal is to get a new handle on certain aspects of the
\rep\ theory of \aff s, in particular to obtain simple formul\ae\ for
the characters of \ihwm s. While there exists a closed expression, the well
known Weyl\hy Kac formula \cite{KAc3}, for all these characters, it is often 
difficult to evaluate because it involves a summation over the Weyl group of 
the \hsa\ \gb\ of the affine algebra \g. Therefore one often prefers to have
formul\ae\ which are better controllable, say in terms of infinite sums or 
products that are easy to handle with algebraic manipulation programs. For
instance, in the case of the classical series of simple \lie s \gb\ one would 
like to have a form of the characters which has a simple functional dependence
on the rank of \gb. 

Simple character formul\ae\ of this type are in particular known for the 
level-one modules of many algebras, owing to the fact that these modules can be 
realized in terms of free bosons
(when \gb\ is simply laced) or free fermions (when \gb\ is \son\ or \sun). 
Now all irreducible modules at arbitrary positive integral level
can be obtained as submodules of tensor products of level-one modules.
Therefore in principle the realization through free fields can also be exploited
at higher level. The problem with this approach
is that in general it is extremely difficult
to identify the irreducible submodules in tensor products. In particular, the 
branching `coefficients' for tensor products of \aff\ modules are not numerical 
constants, but have a functional dependence on (part of) the \csa\ of \g. More 
specifically, when the branching rules are expressed in terms of
the characters of the modules, these branching functions constitute 
characters of the observable algebra of the coset \cft\
  $$ {\cal C}os \,\simeq\,
  (\g_{{\rm level}\;1})^{\oplus\kV} / \g_{{\rm level}\;\kV} $$ 
(see e.g.\ \cite{kawa}). These coset characters are sometimes known even when
the branching rules are not, e.g.\ when
the coset theory can be described as a \cft\ also in a different manner.

Our approach to the problems outlined above is based on the following idea.
We define a field algebra $\fFg$ which is essentially the \cara\ 
of \kv\ species of free fermions
acting on a big Fock space which is the \kv-fold tensor product
of the Fock space of the level-one theory. (To avoid technical complications,
for the purposes of this paper we restrict our attention to the \NS sector of
the fermions.)\,%
\futnote{
Instead of free fermions, one might also use free bosons to implement our
ideas. However, technically these are more difficult to handle because in 
order to deal with genuine conformal fields one must study vertex operators.
In contrast, free fermions are proper conformal fields already themselves.} 
This \alg\ comes with a natural symmetry group \Ok\ (for real
fermions, \resp\ \Uk\ for complex fermions) which, roughly speaking, rotates
the different fermion species into each other.
It is this group \Ok\ (\resp\ \Uk) which we propose as a substitute for the 
gauge group in the DHR sense. Accordingly we introduce an `intermediate'
observable algebra $\fAg$, to which we will refer as the 
{\em gauge invariant fermion \alg}, which is defined as the gauge invariant 
subalgebra of the field algebra $\fFg$. The observables of the level-\kv\ 
WZW theory are naturally gauge invariant, and hence they do not 
make transitions between the sectors of $\fA$. 
Let us denote by $\AW$ the observable algebra of {\em bounded\/}
operators which is associated to the WZW model; it can be
constructed from the positive energy representations of the
corresponding loop group acting on the big Fock space.
The algebra $\AW$ contains the bounded
functions of local current operators, and the irreducible \rep\ spaces of 
the positive energy \rep s are precisely the \hwm s of the chiral symmetry 
algebra of the \wzwt. The latter is given by the semidirect sum of the
untwisted \aff\ \sonh\ at level \kv\ and the Virasoro \alg\ that is 
associated to \sonh\ by the Sugawara construction (hence in particular
it consists of {\em unbounded\/} operators).

Just like in the DHR situation, the sectors of the gauge invariant fermion 
algebra $\fAg$ can be described with the help of the \rep\ theory of the gauge 
group. However, except for level one, these are different from the 
sectors of the \wzwt, because the observables \AW\ of the WZW theory do not
exhaust the invariants of the gauge group, i.e.\ we have the proper inclusions
  $$  \AW \,\subset\, \fAg \,\subset\, \fFg \,. $$ 
Nevertheless a lot of 
information about the decomposition of tensor products of level-one modules
into modules of the level-\kv\ chiral algebra can be obtained by 
decomposing the big Fock space into the irreducible sectors of $\fA$. 
This is possible because the Virasoro algebra of the coset \cft\
$(\g_{{\rm level}\;1})^{\oplus\kV}\! / \g_{{\rm level}\;\kV}$ is
gauge invariant as well, so that we can combine this decomposition with
information on the \rep\ theory of the coset Virasoro algebra. 

In this paper we will examine in detail a non-trivial theory which already
displays the generic features of the superselection structure, but can still 
be managed without having to delve into too many technicalities.
Namely, we will treat the case of two species of real fermions,
corresponding to the gauge group \Oz. The simplicity of this example
can be regarded as reflecting the fact that the \rep s of \Oz\ are at 
most \twodim. Furthermore, information about
this theory is also available from other sources, namely \cite{scya5} certain
conformal embeddings of \aff s, so that we can cross-check our results.
The study of more complicated theories will be left to future work.

Our paper ist organized as follows. In the next three sections we describe 
the algebraic aspects of free fermions: the \cara\ (\srf1), the associated
gauge group \Ok\ (\srf{15}), and some specific features of the level-two 
gauge group \Oz\ (\srf2). Afterwards we provide some basic information about 
the various Lie algebraic and \cft\ structures that will be employed: 
the simple \lie\ \son\ (\srf3), the affine \lie\ \sonh\ and the 
spectrum of the associated \wzwt\ at levels one and two (\srf4), 
and the $\zet_2$-orbifold \cfts\ with conformal central charge $c=1$ (\srf5).
Then we proceed to the analysis of the decomposition of tensor products of 
level-one \sonh-modules. First the \hwv s of \sonh\ at level two and of 
the coset Virasoro algebra (\srf8) are identified. We are then in a position to 
compute the characters of the sectors of $\fA$ (sub\srf9.1) and of
\sonhz\ (sub\srf9.2 and \ref{s.9}.3).
In the final \srf c we summarize our results on the tensor product 
decomposition, and we also remark on implications of the \rep\ theory of 
the gauge group for the fusion rules of the \wzwt.

\Sect{The CAR algebra}1

We consider a separable Hilbert space $\KK$ endowed with an
anti-unitary involution $\GG$ (complex conjugation), $\GG^2=\id$, which obeys
  \be  \langle \GG f, \GG g \rangle = \langle g,f \rangle  \ee
for all $f,g\iN\KK$. The selfdual \car\ (CAR) algebra $\CKG$ corresponding to a
single free fermion
is defined to be the $C^*$-norm closure of the algebra that is generated 
by the range of a linear mapping $B\!:f\mapsto B(f)$ of the Hilbert space
which possesses the property that
  \be  \{ B(f)^*,B(g) \} = \langle f,g \rangle \,{\bf 1}\,,
  \qquad B(f)^*=B(\GG f)  \ee
holds for all $f,g\iN\KK$ (see e.g.\ \cite{arak5}). 

By definition, a quasi-free state $\omega$ of $\CKG$ fulfills 
  \begin{eqnarray}
  && \omega (B(f_1) \cdots B(f_{2n+1})) = 0\,, \\
  && \omega (B(f_1) \cdots B(f_{2n}))= (-1)^{n(n-1)/2} \sum_\sigma
  \mbox{\rm sign}(\sigma)\prod_{j=1}^n \omega
  (B(f_{\sigma(j)})B(f_{\sigma(n+j)}))  \end{eqnarray}
for all $n\iN {\mathbb N}$, where the sum runs over all permutations
$\sigma \iN {\cal S}_{2n}$ with the property
  \be  \sigma (1) < \sigma (2) < \ldots < \sigma (n) 
  \qquad {\rm and}\qquad \sigma(j)<\sigma(j+n)\quad{\rm for}\ j=\oneton \,. \ee
Quasi-free states are completely characterized by
their two point functions. Moreover, the formula
  \be \omega (B(f)^* B(g)) = \langle f,Sg \rangle  \labl{phiS}
provides a one-to-one correspondence between the set of quasi-free states of
$\CKG$ and the subset
  \be  \QKG:= \{S\iN{\mathfrk B}(\KK) \,|\, S=S^*,\, 0 \le S \le {\bf 1},\,
  S + \GG S \GG = {\bf 1} \} \ee
of ${\mathfrk B}(\KK)$ (the set of bounded operators on $\KK$).
It is therefore convenient to denote the quasi-free state 
characterized by equation \erf{phiS} by $\omega_S$. The projections in 
$\QKG$ are called basis projections or polarizations. 
For a basis projection $P$, the state $\omega_P$ 
is pure and is called a Fock state. The corresponding 
GNS representation $({\cal H}_P,\pi_P,|\Omega_P\rangle)$
is irreducible; it is called the Fock representation.
The space ${\cal H}_P$ can be canonically identified
with the antisymmetric Fock space ${\cal F}_{\!-}^{}(P\KK)$. 

Let us now specialize to the Hilbert space
  \be  \KK=L^2(S^1;{\mathbb C}^N)\equiv L^2(S^1)\otimes{\mathbb C}^N \,,\ee
which corresponds to a fermion living on the circle $S^1$ and
carrying the $N$-dimensional vector \rep\ of the \lie\
\son. The involution $\GG$ is given by component-wise
complex conjugation. We introduce a (Fourier) orthonormal basis
  \be  \{e_r^i\mid r\iN\zet+\onehalf,\, i=\onetoN\} \ee
of $\KK$ by the definition
  \be  e_r^i:= e_r\otimes u^i\qquad {\rm for}\quad
  r\iN\zet+\onehalf\,,\ i=\onetoN \,, \ee
where $e_r\iN L^2(S^1)$ are defined by
$e_r(z)=z^r$ (with $z=\eE^{\ii\varphi}$, $-\pi<\varphi\le\pi$), and where
$u^i$ denote the canonical unit vectors of ${\mathbb C}^N$.
The \NS operator $\PNS\iN \QKG$ is then by definition the basis projection
  \be  \PNS:= \sum_{i=1}^N \sum_{r\in\natnumo+\h} 
  |e_{-r}^i \rangle\, \langle e_{-r}^i | \,. \ee
The GNS representation associated to the Fock state $\omega_{\PNS}$
provides the Fock space $\HNS$ which decomposes into the basic
and the vector module of $\sonh$ at level one.
(In this paper we only discuss the \NS sector. The Ramond sector, in which
a Fourier basis with integer powers of $z$ appears 
will not be considered here. 
It could be analyzed by the same methods, but the technical details are
considerably more involved.)

We are interested in the theory that is obtained when one considers an arbitrary
number \kv\ of \NS fermions of the type described above. Thus in addition to the
\son\ index $i$ the fermion modes will now be labelled by a `flavor' index
$q$ which takes values in $\{\qk\}$. To describe this theory, we define
  \be  \KKh:=\KK\otimes\complex^\kV ,\quad \GGh:=\GG\otimes\GG_\kV
  \quad {\rm and}\quad
  \PNSh:=\PNS\otimes \one_\kV\,,  \labl{hat}
where $\GG_\kV$ denotes the canonical complex conjugation in
$\complex^\kV$. Further, for any $f\iN\KK$ we define the elements
  \be  B^q(f):=B(f\otimes v^q)\,, \qquad q=\qk\,,\ee
of $\CKGh$, where $v^q$ denote the canonical unit vectors of $\complex^\kV$.
By $(\HNSh,\PiNSh,\ONSh)$ we denote the GNS representation
associated to the Fock state $\omega_{\PNSh}$ of $\CKGh$; we will refer 
to $\HNSh$ as the `big Fock space'. We then define the Fourier modes
  \be  \b iqr := \PiNSh (B^q(e_r^i)) \labl{biqr}
for $i=\onetoN$, $q=\qk$ and $r\iN{\mathbb Z}+\frac12$. 
The Fourier modes $\b iqr$ generate a CAR algebra with relations
  \be  \{\b ipr, \b jqs \} \, = \, \del pq \, \del ij
  \, \del r{-s} \, {\bf 1}\,.  \ee
The modes $\b iqr$ with positive
index $r$ act as annihilation operators in $\HNSh$, i.e.\ for all $q=\qk$
and all $i=\onetoN$ we have  
  \be  \b iqr\, \ONSh =0 \qquad {\rm for}\ r\in{\mathbb N}_0 +\onehalf \,. \ee

\Sect{The gauge group \Ok}{15}

Field and observable algebras of the fermion theory are described as follows.
Choose a point $\zeta\iN S^1$ on the circle and denote by $\Jzeta$ the
set of those open intervals $I\subseT S^1$ whose closures do not contain
$\zeta$. For $I\iN\Jzeta$ let $\KK (I)$
be the subspace of functions having support in $I$. Correspondingly,
define $\KKh (I)=\KK (I)\otimes\complex^\kV$. The local field 
algebras $\fF (I)$ are then defined to be the von Neumann algebras 
  \be \fF (I) = \PiNSh (\CKGIh)'' \labl{local}
(the prime denotes the commutant in $\mathfrak{B}(\HNSh)$),
and the global field algebra $\fFg$ is the $C^*$-algebra that is defined
as the norm closure of the union of the local algebras,
  \be \fFg= \overline{\bigcup_{I\in\Jzeta} \fF (I)}\,. \ee
Now the group \Ok\ acts in the natural way on the multiplicity
space $\complex^\kV$ in \erf{hat}, and this extends canonically to an action
on $\CKGh$ by Bogoliubov automorphisms.
Moreover, by construction, these automorphisms leave the Fock state 
$\omega_{\PNSh}$ invariant. Hence we obtain a unitary representation $U$
of \Ok\ in $\HNSh$. Also its action respects the local
structure \erf{local}, and thus \Ok\ can be regarded as a substitute for
the gauge group in the sense
of Doplicher, Haag and Roberts \cite{dohr}. This a subgroup of the
automorphism group of $\fF (I)$ \resp\ $\fFg$ such that the observables
are precisely the gauge invariant fields. Therefore the local 
observable algebras $\fA (I)$ and the global (or quasi-local)
observable algebra
$\fAg$ are defined as \Ok-invariant part of the field algebras,
  \be \fA (I) = \fF (I) \cap U(\Ok)' \ee
and
  \be \fAg= \overline{\bigcup_{I\in\Jzeta} \fA (I)}\,. \ee
In the level-one case, the intermediate algebra $\fAg$ coincides with the
observable \alg\ \AW\ of the \wzwt, 
so that the representation theory of $\fAg$ 
reproduces precisely the sectors of the observable algebra \AW; furthermore,
the DHR product of the sectors of the gauge (i.e.\ $\Oe\equiv\zet_2$)
invariant fermion \alg\ $\fAg$ that is obtained by composing localized 
endomorphisms of $\fAg$ provides the WZW fusion rules \cite{bock3}. In 
contrast, at higher level the algebra $\fAg$ no longer coincides with the
observable algebra $\AW$ of the WZW theory. Indeed we will see that already at 
level $\kv=2$ each irreducible $\fAg$-sector is highly reducible 
under the action of the observable algebra \AW. Nevertheless, owing to
$\AW\subset\fAg$ the \rep\ theory of $\fA$ is crucial for our analysis of the
decomposition of the big Fock space into tensor products of
\hwm s of the level-2 chiral algebra and of the coset Virasoro algebra. 

For the construction of the \hwv s 
within the $\fAg$-sectors it is convenient to work with 
the unbounded operators of $\sonh$ (instead of the bounded
elements of $\AW$) and of the Virasoro \alg\ that is
associated to \sonh\ (at fixed level) by the Sugawara formula.
The generators of this Virasoro \alg, i.e.\ the Laurent modes of the
energy-momentum tensor of the \wzwt, will be denoted by $L_m$.
Also, we denote by \LNS m the Laurent components of the canonical 
energy-momentum tensor of the fermion theory in the \NS \rep, i.e.
  \be  \LNS m = \sumqv \,\, L_m^{(q)}  \qquad {\rm with}\quad L_m^{(q)} 
  = -\half \sum_{i=1}^N \sum_{r\in\zet+1/2} 
  \!r\normord{\b iqr\b iq{m-r}} \,. \labl{LNS}
Thus in particular 
  \be L_0^{(q)} = \sum_{i=1}^N \sum_{r\in\natnumo+1/2} \! 
  r\, \b iq{-r}\b iqr \,. \ee

The Bogoliubov automorphisms act as rotations on the `flavor' index $q$ of 
the fermions. As a consequence, they leave expressions of the form
  \be \sumqv \, B^q(f)\, B^q(g)\,\qquad (f,g \in \KK) \ee
invariant. 
In particular, owing to the summation on $q$ in the bilinear expression
\erf{LNS}, the Virasoro generators $\LNS m$ are \Ok-invariant. 
This implies that the coset Virasoro operators 
  \be  L\coset_m:=\LNS m - L_m  \ee
are gauge invariant as well.

\Sect{The gauge group at level 2}2

Let us now specialize to the case $\kv=2$. Thus we consider the situation
  \be  \KKh=\KK\oplus\KK\,,\quad \GGh=\GG\oplus\GG\quad {\rm and}\quad
  \PNSh=\PNS\oplus\PNS\,.  \ee
Then the transformations
  \be \bearl \sgamt (B^1(f)) := \cost B^1(f) - \sint B^2(f) \,, \\[.3em]
  \sgamt (B^2(f)) := \sint B^1(f) + \cost B^2(f) \eear \labl{sgamt}
for $t\iN\reals$ and
  \be  \seta (B^1(f)):= B^1(f)\,,\qquad \seta 
  (B^2(f)):= - B^2(f)\,. \labl{seta}
define Bogoliubov automorphisms of $\CKGh$ generating
the group \Oz. The invariance of the Fock state $\omega_{\PNSh}$ reads now
  \be \omega_{\PNSh} \circ \sgamt = \omega_{\PNSh}
  = \omega_{\PNSh} \circ \seta \,\,, \ee
and there is a unitary (strongly continuous) 
representation $U$ of $O(2)$ by certain implementers
$U(\sgamt),U(\seta)\iN\mathfrak{B}(\HNSh)$ which satisfy
  \be  U(\sgamt)\, \ONSh = \ONSh = U(\seta)\, \ONSh \,, \ee
and the action of $\sgamt$ and $\seta$ extends to $\mathfrak{B}(\HNSh)$. 

The inequivalent \findim\ \irrep s of \Oz\ are the following.
Besides the identity $\rpo$ with $\rpo(\cdot)=1$ and another \onedim\ \rep\
$\rpj$ with
  \be  \rpj(\sgamt)=1\,, \qquad \rpj(\seta)=-1  \,, \labl{rpj}
there are only \twodim\ \rep s $\rp m$ with $m=1,2,...$\,; their \rep\ matrices
are
  \be  \rp m(\sgamt)= \smallmatrix{{cc} \eE^{\ii mt}&0\\0&\eE^{-\ii mt}}
  \,,\qquad \rp m(\seta)= \smallmatrix{{cc} 0&1\\1&0}  \,.  \labl{rp}
The tensor product decompositions of these \rep s read
  \be  \bearl  \rpj\tp\rpj=\rpo\,, \qquad\quad \rpj\tp\rp m=\rp m\,,  \nline8
  \rp m\tp\rp n = \rp{|m-n|} \oplus \rp{m+n} \qquad{\rm for}\ m\ne n\,, \nline6
  \rp n\tp\rp n = \rpo \oplus \rpj \oplus \rp{2n} \,.  \eear \labl{tp}

Employing the results of \cite{dohr}, it then follows that the Hilbert space 
$\HNSh$ decomposes into irreducible sectors of the global observable 
algebra $\fAg$ as
  \be \HNSh = \Hho \oplus \Hhj \oplus
  \bigoplus_{m=1}^\infty (\Hhm \otimes H_{[m]}) \,. \ee
Here $\Hho,\,\Hhj$ and $\Hhm$ carry mutually inequivalent
irreducible representations of $\fAg$; vectors 
in $\Hho,\Hhj$ transform according to the two inequivalent \onedim\ \irrep s
$\rpo$ and $\rpj$ of the gauge group \Oz, respectively, and
the $H_{[m]}\simeq\complex^2$ carry the inequivalent \twodim\ irreducible
\Oz-representations $\rp m$. Later we will also use the notation
  \be  \Hhm\otimes H_{[m]}=\Hhm^+\oplus\Hhm^- \,, \ee
where by definition, $U(\sgamt)$ acts on $\Hhm^\pm$ as
multiplication with $\eE^{\pm\ii mt}$.

\Sect{The simple \lie\ \son}3

Normal-ordered bilinears in the free fermions modes that were introduced in 
\srf1 realize the affine \kma\ \sonh\ at level \kv. To describe this
realization of \sonh, we first need to collect some 
information about the simple \lie\ \son\ which is canonically embedded in \sonh.
We denote by $\el$ the rank of the \lie\ \son, i.e.\ $\el=N/2$ and 
$\el=(N-1)/2$ for even and odd $N$, \resp. 

We use the notation
  \Be  \T ij:=\ii\,(E^{i,j}-E^{j,i})  \Ee
for $i,j=\onetoN$, where $E^{i,j}$ are the matrix units, which have
entries $(E^{i,j})_{k,l}=\del ik\del jl$. The matrices $\T ij$ satisfy
  \be  [\T ij,\T kl]=\ii\,( \del jk \T il + \del il \T jk - \del jl \T ik -
  \del ik \T jl ) \,. \labl{T,T}
Next we introduce the combinations
  \be  H^j:= \T{2j-1}{2j} \quad  {\rm for}\ j=\onetol \labl{Hj}
and
  \be  \EO j\pm := \pm \te j{j+1}\pm\mp \ \ {\rm for}\ j=\onetolme\,, \qquad
  \EO\el\pm:= \left\{ \bearll \pm \te{\el-1}\el\pm\pm & {\rm for}\ N=2\el\,, 
  \nline8  \pm \ttpm\el & {\rm for}\ N=2\el+1 \,,  \eear\right.   \labl{Ej}
where
  \be  \bearl
  \ttee ij:= \half(\eps\T{2i}{2j-1}+\eta\T{2i-1}{2j}) + \halfi
  (\T{2i-1}{2j-1}-\eps\eta\T{2i}{2j}) \,, \nline7
  \tte j:= -\fsqz (\eps\T{2j-1}{2\el+1} - \ii\T{2j}{2\el+1}) \eear \labl{ttee}
for $i,j=\onetol$ and $\eps,\eta=\pm1$.
The matrices \erf{Hj} and \erf{Ej} obey the commutation relations
  \be  
  [H^j,H^k]=0 \,, \qquad [\EO j+,\EO k-]= \del jk\, H^j \,, \qquad 
  [H^j,\EO k\pm]=\pm\,(\alpha^{(k)})^j_{}\,\EO k\pm  
  \labl{sozl}
for $j,k=\onetol$, with structure constants
  \be  \bearl
  (\alpha^{(k)})^j_{}= \del jk-\del j{k+1} \qquad\,{\rm for}\ k=\onetolme\,,
  \nline7 
  (\alpha^{(\el)})^j_{}= \left\{ \bearll
  \del j{\el-1}+\del j\el & {\rm for}\ N=2\el \,, \nline7
  \del j\el & {\rm for}\ N=2\el+1 \,. \eear\right.  \eear \labl{sr}
It is also straightforward to check that the matrices $\EO j\pm$ ($j=\onetol$) 
obey the Serre relations of \son. It follows that
\erf{Hj} and \erf{Ej} constitute a \cwb\ for the defining
matrix realization of \son. The \csa\ is spanned by the $H^j$; the vectors
$\alpha^{(k)}$ are the simple roots of \son, and $\EO k+$ are
the step operators corresponding to these simple roots. The step operators
corresponding to positive roots are then $\te ij+-$ and $\te ij++$
with $1\le i<j\le\el$, and the one corresponding to the highest root $\theta$
is $\te12++$.

According to the explicit expressions \erf{sr} we
are working with an orthonormal basis for the weight space
of \son. The relation with the Dynkin basis is as follows.
For $N=2\el$, the components $\mu^i$, $i=\onetol$, of a weight $\lambda$ in the 
orthonormal basis are related to the weights $\lambda^i$ of $\lambda$ in the
Dynkin basis by
  \be  \mu^i=\dsum_{j=i}^{\el-2} \lambda^j + \half(\lambda^{\el-1}+\lambda^\el)
  \quad{\rm for}\ i=\onetolme\,, \qquad
  \mu^\el=\half(\lambda^{\el-1}-\lambda^\el)\,, \ee
or conversely,
  $\lambda^i=\mu^i-\mu^{i+1}$ for 
  $i=\onetolmz$ and  
  $\lambda^{\el-1}=\mu^{\el-1}+\mu^\el$,  
  $\lambda^\el=\mu^{\el-1}-\mu^\el$.      
For $N=2\el+1$, the analogous relations read
  \be  \mu^i=\dsum_{j=i}^{\el-1} \lambda^j + \half\lambda^\el \ee
for all $i=\onetol$, \resp\
  $\lambda^i=\mu^i-\mu^{i+1}$ for 
  $i=\onetolme$,  
  $\lambda^\el=2\mu^\el$.   
Thus in particular for the fundamental weights $\Lj j$ of \son, defined by
  \be  (\alpha^{(j)},\Lj k) = \left\{ \bearll \half\del k\el & {\rm for}\
  j=\el,\,N=2\el+1 \,, \nline8 \del jk & {\rm else} \,, \eear\right. \ee
the components in the orthonormal basis are
  \be  \hsp{-1.4}
  \Lj j = \left\{ \bearll \!\!(\underbrace{1,1,...\,,1}_{j\;{\rm times}}
  ,0,0,...\,,0)\! & {\rm for}\ j=\onetolmz\,\ {\rm or}\,\ j=\el-1,\, N=2\el+1\,,
  \\{}\\[-.8em]
  \!\half(1,1,...\,,1,1,-1) & {\rm for}\,\ j=\el-1,\, N=2\el \,, \\[.4em]
  \!\half(1,1,...\,,1,1,1) & {\rm for}\,\ j=\el \,. \eear\right. \labl{Lj}
Finally we note that the invariant bilinear form on \son\ is 
  \be  ( \T ij | \T kl ) = \half {\rm tr} \, ( \T ij \T kl )
  = \del ik \del jl -   \del il \del jk \,. \labl{ibf}
In particular, we have
  \Be  (H^i|H^j) = \del ij = (\EO i+|\EO j-) $, $ 
  (\EO i\pm|\EO j\pm) = 0. \Ee

\Sect{The affine \lie\ \sonh}4

Given the fermion modes \erf{biqr}, one defines their normal-ordered bilinears
  \be \Jm ij:=\ii\,\sumq \llb \BX ijqm - \BX jiqm \lrb  \,, \labl J
with
  \be  \BX ijqm:=\half\sum_{r\in\zet+1/2} \normord{\b iqr\b jq{m-r}}  \ee
for $q=1,2$ and $i,j=\onetoN$. One checks by direct computation that
  \be  [\Jm ij,\b kqr]=\ii\,(\del jk\b iq{r+m}-\del ik\b jq{r+m}) \,,  \labl{Jb}
and 
  \be  [\J ijm,\J kln]=\ii\,( \del jk \J il{m+n} 
  + \del il \J jk{m+n} - \del jl \J ik{m+n} - \del ik \J jl{m+n} ) +
  2 \, m \, \del m{-n} \, ( \del ik \del jl - \del il \del jk ) \,. \labl{JJ}
According to \erf{JJ} (compare also \erf{ibf}), the $\Jm ij$ with $i<j$ 
provide a basis for the affine \lie\ \sonh\ at fixed value $\kv=2$ of the level.
That the level of \sonh\ has the value 2 is of course a consequence of the
summation over two species of fermions in \erf J;
for a single fermion one obtains analogously the \lie\ \sonh\ at level 1.
Also note that in the orthonormal basis the \hw s of integrable \hwm s satisfy
  \Be  \mu^0+\mu^1+\mu^2=\kv.  \Ee

A Chevalley basis of the affine \lie\ \sonh\ looks as follows.
The \csa\ generators are
  \be  \HH j:= \Jo{2j-1}{2j}  \labl{HH}
for $j=\onetol$, and the step operators for the simple roots
(\resp\ minus the simple roots) are $\EE j\pm$ with $j=\otol$, given by
  \be  \bearl
  \EE j\pm=\pm\Jot j{j+1}\pm\mp \qquad {\rm for}\ j=\onetolme\,, \nline7 
  \EE0\pm= \pm\Jt{\pm1}12\mp\mp \,, \qquad
  \EE\el\pm = \left\{ \bearll \pm\Jot{\el-1}\el\pm\pm & \forzl \,, \nline7
       \pm J_0(\ttpm \el) & \forzle\,, \eear\right.  \eear \ee
where 
  \be  \bearl
  \Jmtee ij:= \half(\eps\Jm{2i}{2j-1}+\eta\Jm{2i-1}{2j}) + \halfi
  (\Jm{2i-1}{2j-1}-\eps\eta\Jm{2i}{2j}) \,, \nline7 
  \Jmte j:= -\fsqz (\eps\Jm{2j-1}{2\el+1} - \ii \Jm{2j}{2\el+1})  
  \eear \labl{Jmte}
for $i,j=\onetol$ and $\eps,\eta=\pm1$.

At any integral level \kv\ the \aff\ \sonh\ has a finite number of \ihwm s
$\Hhkvl$; in the \aft\ description they correspond to the positive energy
\rep s of the loop group $\mathit{LSO}(N)$ 
that are carried by the \suse s. For 
level one and level two these are listed in the tables \ref{T1} -- \ref{t2}.
In these tables, $\Lambda$ denotes the \hw\ \wrtt horizontal subalgebra \son, 
$\Delta$ the conformal weight, and \qdi\ the 
statistical (or quantum) dimension.
In the first column we provide a `name' for the associated primary field of the
relevant \wzwt; below we will use these names as labels for the \ihwm s, i.e.\
write $\Hhzl=\Hhzo$ for $\Lambda=0$ etc., and for other quantities such as
characters. (We find it convenient to use identical names for some of the 
fields at level one and at level two; when required to avoid ambiguities in 
the notation, we will always also specify the level.)

\begin{table}[bpth]\caption{Irreducible \hwm s of \sonh\ at level 1 
for $N=2\el$ (left) and for $N=2\el+1$ (right).} \label{T1}
\begin{center}
  \begin{tabular}{|c|c|c|c|} \hline &&&\\[-.9em]
  field & $\Lambda$   & $\Delta$      & \qdi \\ \hline\hline &&&\\[-.9em]
  $\rmo$&   0         & 0             & 1    \\ &&&\\[-.9em]
  \rmv  & $\Lj1$      & $\Frac12$     & 1    \\ &&&\\[-.8em] \hline &&&\\[-.9em]
  \rms  & $\Lj{\el-1}$& $\Frac N{16}$ & 1    \\ &&&\\[-.8em]
  \rmc  & $\Lj\el$    & $\Frac N{16}$ & 1    \\[-.8em]&&&\\ \hline \end{tabular}
  \hsp5
  \begin{tabular}{|c|c|c|c|} \hline &&&\\[-.9em]
  field & $\Lambda$   & $\Delta$      & \qdi \\ \hline\hline &&&\\[-.9em]
  $\rmo$&   0         & 0             & 1    \\ &&&\\[-.9em]
  \rmv  & $\Lj1$      & $\Frac12$     & 1    \\ &&&\\[-.8em] \hline &&&\\[-.9em]
  $\sigma$ & $\Lj\el$    & $\Frac N{16}$ & $\sqrt2$
  \\[-.8em]&&&\\ \hline \multicolumn4c {} \\[.5em] \end{tabular}
\end{center} \end{table}

\begin{table}[ptbh]\caption{Irreducible \hwm s of \sonh\ at level 2 
for $N=2\el$.} \label{t1}
\begin{center}
  \begin{tabular}{|c|c|c|c|} \hline &&&\\[-.9em]
  field & $\Lambda$    & $\Delta$     & \qdi \\ \hline\hline &&&\\[-.9em]
  $\rmo$&   0          & 0            & 1    \\ &&&\\[-.9em]
  \rmv  & $2\Lj1$      & 1            & 1    \\ &&&\\[-.8em]
  \rms  & $2\Lj{\el-1}$& $\Frac N8$   & 1    \\ &&&\\[-.8em] 
  \rmc  & $2\Lj\el$    & $\Frac N8$   & 1    \\ &&&\\[-.8em]
  $j$   & $\left\{ \bearl \Lj j\ \;{\rm for}\ j=\onetolmz, \\[.1em]
          \Lj{\el-1}+\Lj\el\ {\rm for}\ j=\el-1 \eear\right.$
                 & $\Frac{j(N-j)}{2N}$ & 2  \\ &&&\\[-.8em] \hline &&&\\[-.9em]
  $\sigma$ & $\Lj{\el-1}$ & $\Frac{N-1}{16}$ & $\sqrt{N/2}$   \\ &&&\\[-.8em]
  $\tau$   & $\Lj\el$     & $\Frac{N-1}{16}$ & $\sqrt{N/2}$   \\ &&&\\[-.8em]
  $\sigma'$& $\Lj1+\Lj{\el-1}$ & $\Frac{N+7}{16}$ & $\sqrt{N/2}$ \\ &&&\\[-.8em]
  $\tau'$  & $\Lj1+\Lj\el$     & $\Frac{N+7}{16}$ & $\sqrt{N/2}$ 
  \\[-.8em]&&&\\ \hline \end{tabular}
\end{center} \end{table}

\begin{table}[ptbh]\caption{Irreducible \hwm s of \sonh\ at level 2 
for $N=2\el+1$.} \label{t2}
\begin{center}
  \begin{tabular}{|c|c|c|c|} \hline &&&\\[-.9em]
  field & $\Lambda$    & $\Delta$     & \qdi \\ \hline\hline &&&\\[-.9em]
  $\rmo$&   0          & 0            & 1    \\ &&&\\[-.9em]
  \rmv  & $2\Lj1$      & 1            & 1    \\ &&&\\[-.8em]
  $j$   & $\left\{ \bearl \Lj j\ \;{\rm for}\ j=\onetolme, \\[.1em]
                  2\Lj\el\ {\rm for}\ j=\el \eear\right.$
                 & $\Frac{j(N-j)}{2N}$ & 2  \\ &&&\\[-.8em] \hline &&&\\[-.9em]
  $\sigma$ & $\Lj{\el-1}$ & $\Frac{N-1}{16}$ & $\sqrt{(N-1)/2}$ \\ &&&\\[-.8em]
  $\sigma'$& $\Lj1+\Lj{\el-1}$ & $\Frac{N+7}{16}$ & $\sqrt{(N-1)/2}$ 
  \\[-.8em]&&&\\ \hline \end{tabular}
\end{center} \end{table}
  
In the tables we have separated the modules by a horizontal line into two 
classes. In the fermionic description, the modules in the first part are in 
the \NS sector, while those in the second part are in the Ramond sector.
As we only treat the \NS sector of the fermions here, we will not deal with
the second class of \rep s; we have included them in the tables only for
completeness. Thus at level one we have $\HNS=\Hheo\oplus\Hhev$, and hence 
at level two we can write
  \be \HNSh = (\Hheo\otimes\Hheo) \oplus (\Hheo\otimes\Hhev)
  \oplus (\Hhev\otimes\Hheo) \oplus (\Hhev\otimes\Hhev) \,.  \labl{HNSh}
The four summands in this decomposition can be characterized as the common 
eigenspaces \wrt the `fermion flips' $U(\sgam_\pi \seta)$ and $U(\seta)$,
namely those associated to the pairs $(1,1)$, $(1,-1)$, $(-1,1)$ and $(-1,-1)$ 
of eigenvalues, \resp. By comparison with the action \erf{rpj} and \erf{rp}
of \Oz\ on the $\fA$ sectors, it follows that we can decompose the tensor 
products appearing in \erf{HNSh} as
  \be \bearl
  \Hheo\otimes\Hheo = 
  \Hho \oplus \dstyle\bigoplus_{n=1}^\infty \Hh {[2n]} \,, \qquad 
  \Hhev\otimes\Hhev = 
  \Hhj \oplus \dstyle\bigoplus_{n=1}^\infty \Hh {[2n]} \,,\\{}\\[-.8em]
  \Hheo\otimes\Hhev = \dstyle\bigoplus_{n=0}^\infty 
  \Hh {[2n+1]} = \Hhev\otimes\Hheo \,.  \eear \labl{Hdec}

The (Virasoro specialized) character of an \ihwm\ is the trace of
$q^{L_0}$ over the module, where $L_0$ is the zero mode of the \emt\ (see 
\erf{LNS}) and where $q$ is either regarded as a formal variable, or as
$q=\exp(2\pi\ii\tau)$ with $\tau$ in the upper complex half plane.%
\,\futnote{To obtain simpler transformation behavior
\wrt modular transformations of the variable $\tau$, one often defines the
character with an additional factor of $q^{-c/24}$. For our purposes, this
modification is not needed.}
The characters of the modules in the \NS sector at level one are
  \be  \chieo(q)= \frac{(\vi(-q^\h))^N   
  + (\vi(q^\h))^N}{2\,(\vi(q))^N}\,,  \qquad
  \chiev(q)= \frac{(\vi(-q^\h))^N        
  - (\vi(q^\h))^N}{2\,(\vi(q))^N}\,,  \ee
where 
  \be  \vi(q):= \prod_{n=1}^\infty (1-q^n) \, \labl{vi}
is Euler's product function.

In \srf9 we will employ the \rep\ theory of the gauge group \Oz, and in
particular the decomposition \erf{Hdec}, to obtain also simple formul\ae\ 
for the characters of the level-two modules in the \NS sector. As further
input, we will need some information about the relevant coset \cfts.

\Sect{$c=1$ orbifolds}5

Via the coset construction \cite{goko2}, one associates to any embedding of
untwisted \aff s that is induced by an embedding of their \hsa s another
\cft, called the coset theory. Here the relevant embedding is that of
\sonhz\ into $\sonhe\oplus\sonhe$; the branching rules of this embedding
are just the tensor product decompositions of \sonhe-modules.

The Virasoro algebra of the coset theory is easily
obtained as the difference of the Sugawara constructions of the 
Virasoro algebras of the \aff s. In contrast, the determination of the 
field contents of the coset theory is in general a different task (see e.g.\ 
\cite{scya6,fusS4}). But in the case of our interest, the coset theory has
conformal central charge $c=1$, and the classification of (unitary) $c=1$
\cfts\ is well known. In fact, one finds 
(compare e.g.\ \cite{scya5}) that it is a so-called rational
$c=1$ orbifold theory, which can be obtained from the $c=1$ theory of a
free boson compactified on a circle by restriction to the invariants \wrt
a $\zet_2$-symmetry. These \cft\ models have been investigated in \cite{dvvv};
for our purposes we need only the following information.

The rational $c=1$ $\zet_2$-orbifolds are labelled by a non-negative integer 
$M$. The theory at a given value of $M$ has $M+7$ sectors; they are listed in
the following table.
  \be
  \begin{tabular}{|c|c|c|} \hline &&\\[-.9em]
  field         & $\Delta$         & \qdi      \\ \hline\hline &&\\[-.9em]
  $\rmo$        & 0                & 1         \\ &&\\[-.9em]
  \rmv          & 1                & 1         \\ &&\\[-.8em]
  \rms, \rmc    & $\Frac M4$       & 1         \\ &&\\[-.8em]
  $j\ \ \iN\{\onetomme\}$ &$\Frac{j^2}{4M}$ &2 \\ &&\\[-.8em]\hline&&\\[-.9em]
  $\sigma,\,\tau$ & $\Frac1{16}$ & $\sqrt M$   \\ &&\\[-.8em]
  $\sigma',\,\tau'$ & $\Frac9{16}$ & $\sqrt M$ \\[-.8em]&&\\ \hline
  \end{tabular}
  \labl{t3}
Here we have again separated the fields which correspond to the \NS sector
from the fields $\sigma,\,\tau,\, \sigma',\,\tau'$ which correspond to the
Ramond sector; the latter are known as `twist fields' of the orbifold theory.

The characters of the fields in the \NS sector are given by
  \be  \chiCj  (q) = \Frac1{\vi(q)}\, \psim j(q) \labl{chiCj}
for $j=\onetom$, where it is understood that
  \be  \chiCs  (q) = \chiCc  (q) = \half \chiCm \,,\labl{chiCs}
and by
  \be  
  \chiCo  (q) = \Frac1{2\vi(q)}\,\llb \psim 0(q) 
  + \psiemq \lrb\,, \qquad 
  \chiCv  (q) = \Frac1{2\vi(q)}\,\llb \psim 0(q) - \psiemq \lrb\,.  
  \ee
Here the functions $\psim j$ are the infinite sums
  \be  \psim j(q):= \sum_{m\in\zet} q_{}^{(j+2mM)^2/4M} \,. \labl{psim} 
One has \cite[p.\,240]{KAc3}
  \be  \psiemq = \summZ(-1)^m\,q^{m^2} = \frac{(\vi(q))^2}{\vi(q^2)} \,. 
  \labl{76}
It follows in particular that
  \be   \chiCo(q) - \chiCv(q) = \frac{\vi(q)}{\vi(q^2)} \,, \labl{chiCovm} 
and
  \be   \chiCo(q) + \chiCv(q) = \frac{\psim 0(q)}{\vi(q)} \,. \labl{chiCovp}

Note that the spectrum of \wzwts\ for even and odd $N$, displayed in tables
\ref{T1} -- \ref{t2}, is rather similar. However, to obtain the spectrum of
the coset theory also the structure of the conjugacy classes of \son-modules
play an important \role, and these are rather different for even and odd $N$.%
\,\futnote{Also, for odd $N$ in the Ramond sector an additional complication 
arises, namely a so-called fixed point resolution is required
\cite{scya5,fusS4}.}
As a consequence it depends on whether $N$ is even or odd which $c=1$ orbifold
one obtains as the coset theory. Namely, for $N=2\el$ one finds $M=N/2=\el$, 
while $M=2N$ for $N=2\el+1$.

The decomposition of the products of level one characters looks as follows.
For $N=2\el$ we have
  \be \bearl 
  [\chieo ]^2 = \chicol \, \chizo  + \chicvl \, \chizv  
    + \dstyle\sum_{2\le j \le \el \atop j \,\,{\rm even}} \chicjl 
  \, \chizj \,,   \nline7 
  [\chiev ]^2 = \chicol \, \chizv  + \chicvl \, \chizo
    + \dstyle\sum_{2\le j \le \el \atop j \,\,{\rm even}} \chicjl 
  \, \chizj\,,   \nline7 
  \chieo \, \chiev  
  = \dstyle\sum_{1\le j \le \el \atop j \,\,{\rm odd}} \chicjl \, 
  \chizj  \,, \eear \labl{anse}
where it is understood that
  \be  \chizl (q) \equiv \chizs (q) + \chizc (q) \,.  \hsp{6.9} \labl{chizs}
For $N=2\el+1$, the tensor product decomposition reads instead
  \be  \hsp3 \bearl
  [\chieo]^2 = \chicozN \, \chizo + \chicvzN\, \chizv + 
  \dstyle\sum_{2\le j \le \el \atop j \,\,{\rm even}}\! \chicjzN {2j}  
  \, \chizj  + \dstyle\sum_{1\le j \le \el \atop j \,\,{\rm odd}}\! 
  \chicjzN{2N-2j}  \, \chizj  \,, \nline7 
  [\chiev ]^2 = \chicozN \, \chizv + \chicvzN\, \chizo + 
  \dstyle\sum_{2\le j \le \el \atop j \,\,{\rm even}}\! \chicjzN {2j}  
  \, \chizj  + \dstyle\sum_{1\le j \le \el \atop j \,\,{\rm odd}}\! 
  \chicjzN{2N-2j}  \, \chizj  \,, \nline7 
  \chieo\, \chiev = \chicjzN {2N}  \,\llb \chizo  + \chizv  \lrb
  \, + \dstyle\sum_{2\le j \le \el \atop j \,\,{\rm even}}\!
  \chicjzN{2N-2j}\, \chizj  + \dstyle\sum_{1\le j\le\el\atop j\,\,
  {\rm odd}}\! \chicjzN {2j}    \, \chizj  \,.  \eear \labl{anso}

It is worth noting that these formul\ae\ can be proven without too much effort,
whereas in general it is a difficult task to write down such 
tensor product decompositions. Tools which are always available are the matching
of conformal dimensions modulo integers as well as conjugacy class
selection rules, which imply \cite{scya6} so-called field identifications.
In the present case, we can e.g.\ use the fact that the sum of conformal weights
$\Delta\kz_j=j(N-j)/2N$ and $\Delta\Coset_k=k^2/4M$ is (for generic $N$)
\futnot{also: `analyticity in $N$'}
a half-integer only if $k=j\sqrt{2M/N}$ or $k=(N-j)\sqrt{2M/N}$. Also,
there is a conjugacy class selection rule which implies that the tensor
product of modules in the \NS sector yields only modules which are again in
the \NS sector, and the corresponding field identification tells us e.g.\ that
the branching function $b^{{\rm c};\scriptstyle M}_{\rmv,\rmv;\rmv}(q)$
coincides with $b^{{\rm c};\scriptstyle M}_{\circ,\circ;\circ}(q)=\chiCoM(q)$.

As it turns out, we are even in the fortunate situation that together with the
known classification of unitary $c=1$ \cfts, these informations
already determine the tensor product decompositions almost completely. 
In particular, the value of $M$ of the $c=1$ orbifold is determined uniquely, 
and one can prove that there aren't any further field identifications besides 
the ones implied by conjugacy class selection rules. 
The remaining ambiguities can be resolved by checking various
consistency relations which follow from the arguments that we will give in
\srf9 below. Another possibility to deduce \erf{anse} and \erf{anso} is to
employ the conformal embedding of \sonhz\ into \unh\ at level one \cite{scya5},
which corresponds to regarding the real fermions as real and imaginary parts
of complex-valued fermions.

\Sect{Highest weight vectors of \sonhz}8

\subsection{Definition of the vectors}

A \hwv\ $\Phila$ of $\sonhz$ with \hw\ $\Lambda$ is characterized by the
following properties. First, it is annihilated by the step 
operators associated to the horizontal positive roots,
i.e.~for $1\le i<j\le\el$ and $\eps=\pm1$ one has
  \be  \Jot ij+\eps\,\Phila = 0\,, \qquad \mbox{and also } \quad
  J_0 (\ttp k) \, \Phila = 0\;\ \forzle \,;\labl{0hor}
second, it is also annihilated by the step operators with positive grade,
i.e.~for $m>0$, $i,j=\onetol$ and $\eps,\eta=\pm1$ it satisfies
  \be  \Jmt ij\eps\eta\,\Phila = 0\,,  \qquad \mbox{and also } \quad
  J_m (\tte k) \, \Phila = 0\;\ \forzle \,; \labl{0pos}
and third, $\Phila$ is an eigenvector of the \csa,
  \be \HH k \, \Phila = \Lambda^k \, \Phila  \labl{Cart}
for $k=\onetol$. 

We will exploit the decomposition of $\HNSh$ into irreducible $\fA$ sectors 
to identify the \hwv s of $\sonhz$. Indeed, in each sector $\Hho$, $\Hhj$ and
$\Hhm^\pm$ we find distinguished states which are \hwv s for both 
$\sonhz$ and the coset Virasoro algebra. The construction works as follows.
The vector $\OM\equiv\ONSh$ is a \hws\ of $\g=\sonhz$ with \hw\ zero.
To describe more \hwv s, it is helpful to introduce some notation. We define
  \be   \xpm jr := \fsqz (\cp jr \pm \ii\cbp jr) \,, \qquad
  \xbpm jr :=   \fsqz (\cm jr \pm \ii\cbm jr) \,, \ee
for $j=\onetol$, where
  \be  \cpm jr := \fsqz (\b{2j}1r\pm\ii\b{2j-1}1r) \,, \qquad 
  \cbpm jr := \fsqz (\b{2j}2r\pm\ii\b{2j-1}2r) \,, \labl{cd}
and also, $\forzle$,
  \be  \xbpm{\el+1}r := \fsqz ( \b{2\el+1}1r \pm
  \ii \b{2\el+1}2r ) \,. \ee
Further, we set
  \be  \bearll  \Xpm jr:=\xpm jr\,\xpm{j-1}r\cdots \xpm1r \qquad
  & {\rm for}\ j=\onetol\,, \nline7 
  \Xbpm jr:=\xbpm{j+1}r\,\xbpm{j+2}r\cdots \xbpm\el r \quad
  & {\rm for}\ j=\otolme \,, \eear\ee
and $\Xbpm \el r:=\bfe$.

For any $n=0,1,2,...$\, we can now define the following vectors: We set
  \be  \Omnpm{j}n := \Xpm j\mnh\,\Oonpm n \,,\qquad 
  \mbox{for} \,\,j=\onetol \,, \labl{omnlj}
as well as
  \be  \Ombnpm jn := \left\{ \bearll \Xbpm j\mnh \Xpm\el\mnh \,
  \Oonpm n & \forzl  \,,   \,\, j=\onetolme \,,
  \nline6    \Xbpm j\mnh \xbpm{\el+1}\mnh \Xpm\el\mnh \,
  \Oonpm n & \forzle \,, \,\, j=\onetol \,.  \eear\right. \labl{ombmlj}   
Here we defined recursively
  \be \Oonpm{n+1} := \left\{\bearll \Xbpm0\mnh \Xpm\el\mnh\, \Oonpm n & \forzl
  \,, \nline6    \Xbpm 0\mnh \xbpm{\el+1}\mnh \Xpm\el\mnh \,
  \Oonpm n & \forzle \,, \eear\right. \labl{Ombmlj}   
with
  \be  \Oonpm 0 := \OM  \hsp6 \,. \ee
Further, we set
  \be  \Ovvo \equiv \pm \Ovo := \x 1\mh \y 1\mh\,\OM  \,, \quad
  \Ovnpm n := \xpm 1\mnh \xmp 1\nmh\,\Oonpm n \,, \ \, n=1,2,...\,, \labl{omnv}
and, $\forzl$,
  \be  \Osnpm n := \Omnpm\el n \,,  \qquad
       \Ocnpm n := \xbpm \el\mnh \xbmp\el\nh\,\Osnpm{n} \,.\labl{oi}

\subsection{\Oz\ transformation properties}

The vacuum $\OM$ is \Oz-invariant. We then deduce (compare the
appendix, \erf{Oz}\,--\,\erf{OZ}) the following
transformations for the vectors \erf{omnlj}\,--\,\erf{oi}. For all
$n=0,1,2,...\,$ we have
  \be  U(\sgamt)\,\Omnpm jn=\eE^{\pm\ii(nN+j)t}\,\Omnpm jn\,,\qquad\hsp{.9}
  U(\seta)\,\Omnpm jn=\Omnmp jn  \labl{sgOmnpm}
for $j=\onetol$, and
  \be  U(\sgamt)\,\Ombnpm jn=\eE^{\pm\ii((n+1)N-j)t}\,\Ombnpm jn\,,\qquad
  U(\seta)\,\Ombnpm jn=\Ombnmp jn  \ee
for $j=\onetol$. Also
  \be  \bearll
  U(\sgamt)\,\Oonpm n=\eE^{\pm\ii nNt} \,\Oonpm n\,,\qquad\hsp{.5}
  & U(\seta)\,\Oonpm n=\Oonmp n \,, \\[.5em]
  U(\sgamt)\,\Ovnpm n=\eE^{\pm\ii nNt} \,\Ovnpm n\,,\qquad\hsp{.5}
  & U(\seta)\,\Ovnpm n=\Ovnmp n \,, \\[.5em]
  U(\sgamt)\,\Osnpm n=\eE^{\pm\ii(nN+\el)t}\,\Osnpm n\,,
  & U(\seta)\,\Osnpm n=\Osnmp n \,, \\[.5em]
  U(\sgamt)\,\Ocnpm n=\eE^{\pm\ii(nN+\el)t}\,\Ocnpm n\,,
  & U(\seta)\,\Ocnpm n=\Ocnmp n \,. \eear  \hsp{.8} \ee

We remark that the \hws s $\Ovnpm n$ and $\Oonpm n$, $n=1,2,\ldots$\,, and for
even $N$ also $\Ocnpm n$ and $\Osnpm n$, $n=0,1,2,\ldots$\,, are connected by 
\Oz-invariant fermion bilinears, i.e.\ by
elements of the intermediate algebra $\fAg$. Explicitly, we have 
  \be  \Ovnpm n = a^n_{\rm v}\, \Oonpm n, \qquad
  a^n_{\rm v} = - (\y1\nmh \x1\mnh + \x1\nmh \y1\mnh)\,, \ee
for $n=1,2,\ldots$, and 
  \be  \Ocnpm n = a^n_{\rm c}\, \Osnpm n, \qquad
  a^n_{\rm c} = - (\yb\el\nh \xb\el\mnh + \xb\el\nh \yb\el\mnh) \labl{sgOO}
for $n=0,1,2,\ldots$.

\subsection{The highest \sonhz\ weights}

The states defined above are eigenvectors of all \csa\ generators
$\HH k$ ($k=\onetol$) and of the central generator $K$; the level \kv\ is
equal to 2, and the weights do not depend on the label $n$. More precisely, 
from the commutation relations \erf{Hx} and \erf{Hxl} it follows rather 
directly that 
  \be  \bearll  \HH k\,\Omnpm jn= (\lj j)^k\,\Omnpm jn 
  & {\rm for}\ j=\onetol\,,
  \\[.5em] \HH k\,\Ombnpm jn = (\lj j)^k\,\Ombnpm jn   
  & {\rm for}\ j=\onetolme \eear \hsp{4.4}\labl{hhk}
and
  \be  \bearll  
  \HH k\,\Oonpm n =(\lo)^k\,\Oonpm n \,, \hsp{2.2} &
  \HH k\,\Ovnpm n =(\lv)^k\,\Ovnpm n \,, \nline7 
  \HH k\,\Osnpm n =(\ls)^k\,\Osnpm n \,, &
  \HH k\,\Ocnpm n =(\lc)^k\,\Ocnpm n \,. \eear \ee
The weights $\lj j$ appearing here are those listed in
table \ref{t1} and \ref{t2}, i.e.\ we have
  \be  \lj j = \left\{ \bearll
  \Lj j & {\rm for}\ j=\onetolmz\ \;{\rm or}\;\ j=\el-1,\;N=2\el+1 \,,\nline7
  \Lj{\el-1}+\Lj\el & {\rm for}\ j=\el-1,\;N=2\el \,,\nline7 
  2\Lj\el           & {\rm for}\ j=\el,\;N=2\el+1 \,, \eear \right. \labl{lj}
with the fundamental weights $\Lj i$ as defined in \erf{Lj}, while
$\lo=0$, $\lv=2\Lj1$, and, $\forzl$, $\ls=2\Lj\el$, $\lc=2\Lj{\el-1}$.

\subsection{The \hw\ property}

Having obtained \erf{hhk}, for proving that the states 
\erf{omnlj}\,--\,\erf{oi} are \hwv s \wrt \sonhz\ it is now
sufficient to show that they are annihilated by $\EE j+$ for $j=\otol$. This 
can easily be checked by inserting 
the results \erf{EX}\,--\,\erf{E0X} for the commutators between the step 
operators $\EE j+$ and the operators $\Xpm kr$, $\Xbpm kr$ into
the definitions of these states.
The least trivial case occurs for $\EE0+$, where one employs the first of the
identities \erf{E0X}; one then has to commute $\xbpm1\h$ and $\xbpm2\h$,
to the right and use $\xbpm1\h\OM=0=\xbpm2\h\OM$ when $n=0$, 
while for $n>0$ one also must employ the second identity in \erf{E0X}.

Thus all the states \erf{omnlj} -- \erf{oi} are \hws s of $\g=\sonhz$.
We claim further that they are \hwv s \wrtt coset Virasoro algebra, too.
This follows directly from the fact that $\LNS m$ with $m>0$ annihilates these 
states, which is a consequence of 
  \be [ \LNS m ,\xpm jr ] = - (r+\Frac{m}2) \,\xpm j{r+m} \,,\qquad
  [ \LNS m ,\xbpm jr ] = - (r+\Frac{m}2) \,\xbpm j{r+m} \,.\labl{Lmx}

Since the affine \lie\ \sonhz\ and the coset Virasoro \alg\ commute, it
follows immediately that further \hwv s of \sonhz\ are obtained when acting
with the creation operators of the coset Virasoro \alg\ on the vectors
\erf{omnlj}\,--\,\erf{oi}. For example,
applying the coset Virasoro operator $L_{-1}^{\rm c}$ to the 
\hwv\ $\Omp 1$ we get the \hwv\ (computed for the case $N=2 \el$)
  \be  L_{-1}^{\rm c} \Omp 1 = \Frac1N\, \LLb \x1{-3/2} \OM + \sum_{k=1}^{\el}
  (\xb k\mh \y k\mh - \yb k\mh \x k\mh) \Omp 1 \LRb   \ee
of \sonhz. However, it follows from the results in \srf9 below that, except
for a few special cases, the vectors \erf{omnlj}\,--\,\erf{oi} exhaust the 
set of simultaneous \hws s of \sonhz\ and the coset Virasoro \alg.

Also note that by construction the tensor product module, and hence each of
its submodules, is unitary. Thus in particular the \hwm s that are obtained
by acting with arbitrary polynomials in the lowering operators $\EE i-$ 
on the \hwv s are unitary, and hence are fully reducible.

\subsection{Conformal weights}

The action of the zero mode of the free fermion Virasoro algebra \erf{LNS}
on the fermion modes $\xpm ir$ reads
  \be [\LNS 0,\xpm ir ] = -r \, \xpm ir \,, \qquad 
  [\LNS 0,\xbpm ir ] = -r \, \xbpm ir \,. \labl{Lx}
{}From these relations we deduce that
  \be \LNS 0 \, \Omnpm jn = \Delns j \, \Omnpm jn \,, \ee
with conformal weights
  \be \Delns j = \big[ \half + \Frac32 + \ldots + \lLb n-\half \lRb
  \big] N + \lLb n+\half \lRb j = \Frac{n^2N}2 + \lLb n+\half \lRb j \ee
for $j=\onetol$. Similarly,
  \be \LNS 0 \, \Ombnpm jn = \Delnsb j \, \Ombnpm jn \,, \qquad
  \Delnsb j = \Frac{(n+1)^2N}2 - \lLb n+\half \lRb j \,,\ee
for $j=\onetol$. Also, for the sectors labelled by $\circ,\,\rmv,\,\rms$
and \rmc\ we find
  \be \Delns \circ = \Frac{n^2N}2 \,,\qquad
  \Delns \rmv = \Frac{n^2N}2+1\,,\qquad
  \Delns \rms = \Delns \rmc = \Delns \el \,. \ee

Furthermore, the conformal weights of the vectors \erf{omnlj}\,--\,\erf{oi}
\wrtt Virasoro algebra of the level-two \wzwt\
follow immediately from the \son-weights $\Lambda$ by 
the Sugawara formula for the Virasoro generator $L_0$.
This yields the conformal weights that were already listed in the
tables \ref{t1} and \ref{t2}. Comparing these conformal dimensions with
the ones obtained above, we arrive at the result 
  \be \bearll\Delc j = \Frac1{2N} (nN+j)^2 \,,\qquad & j=\onetol\,, \nline6 
  \Delcb j = \Frac1{2N} ((n+1)N-j)^2 \,, & j=\onetol\,, 
  \eear \hsp{.9} \labl{ccwa}
and
  \be \Delc \circ = \Delc \rmv = \Frac{n^2N}2 \,,\qquad
  \Delc \rms = \Delc \rmc = \Delc \el  \labl{ccwb}
for the eigenvalues
of the coset Virasoro generator $L\coset_0=\LNS 0-L_0$. 

\Sect{Characters of the modules in the \NS sector}9

Because of the inclusion $\AW\subset\fAg$, the irreducible sectors of
the gauge invariant fermion algebra $\fAg$ constitute modules of the 
observable \alg\ \AW\ of the \wzwt, which however are typically reducible.
To determine the decomposition of the \irmod s of the intermediate \alg\ 
$\fAg$ into \irmod s of \AW\ we analyze their characters and combine the
result with the knowledge about the characters of the coset theory.

\subsection{Characters for the sectors of $\fA$}

The characters of submodules of the space $\HNSh$, i.e.\ the trace of
$q^{L_0}$ over the modules, can be obtained as follows.
Let $\Po, \PJ$ and $\Ppm m$ denote the projections onto $\Hho, \Hhj$
and $\Hhm^\pm$ for $m\iN {\mathbb N}$, respectively.
Then the \rep\ matrices \Ugt\ and \Uget\ of \Oz\ decompose into projectors as
  \be  \Ugt=\Po+\PJ+\sum_{m=1}^\infty \llb 
  \eE^{\ii mt}\Pp m+\eE^{-\ii mt}\Pm m \lrb  \hsp{5.1} \labl{Ugt}
and
  \be  \Uget = \Po - \PJ +\sum_{m=1}^\infty \llb \eE^{\ii mt} 
  U(\seta) \Pp m   +\eE^{-\ii mt}U(\seta) \Pm m \lrb \,.  \labl{Uget}
It follows in particular that the projectors can be written as
  \be  \bearl  \Po = \IntT \llb \Ugt+\Uget \lrb \,, \qquad\ 
  \PJ = \IntT \llb \Ugt-\Uget \lrb \,, \\{}\\[-.4em] 
  \Ppm m = \Intt \eE^{\mp\ii mt} \, \Ugt \qquad{\rm for}\ m\iN {\mathbb N}
  \,.  \eear \labl{Proj}
  
For the irreducible $\fA$-sectors in $\HNSh$, the results \erf{sgOmnpm} --
\erf{sgOO} together with the action of \LNS 0 (compare \erf{Lx}) imply the
following. First,
  \be  \bearll  \chio(q) \!\!& \equiv \trNS \Po\,\qlo  \\[.5em]
  & = \dstyle\intT \llb \prod_{m=0}^\infty (1+\eit\qmh)^N_{}(1+\emit\qmh)^N_{} 
  + \dstyle\prod_{m=0}^\infty (1-\qzme)^N_{} \lrb  \,. \eear \ee
This can be rewritten as
   \be  \chio(q) = \intT \LLb \frac{\wi(q;-\eit q^{1/2})}{\vi(q)} \LRb^N_{}
   + \Frac12\, \LLb \frac{\vi(q)}{\vi(q^2)} \LRb^N_{} \,, \ee
where $\vi$ is Euler's product function \erf{vi} and
  \be  \wi(q;z):= \prod_{n=1}^\infty \lLb (1-q^n)(1-q^nz^{-1})(1-q^{n-1}z) 
  \lRb \,. \labl{wi}
Using also the identity 
  \Be  \wi(q;z) = \sum_{n\in\zet} (-1)^n\,q^{n(n-1)/2}z^n \Ee
\cite[p.\,240]{KAc3}, we finally arrive at 
   \be  \chio(q) = \frac{\sN 0}{2\,(\vi(q))^N}
   + \frac{(\vi(q))^N}{2\,(\vi(q^2))^N} \,, \labl{cnso}
where we introduced the functions
   \be \sN m := \!\! \summN m q_{}^{(m_1^2+m_2^2+...+m_N^2)/2}
   \;\equiv  \sumMN m q_{}^{\vecm^2/2} \, \labl S
for $m\iN\zet$. 

Analogously, we find
   \be  \chiJ(q) \equiv \trNS \PJ\,\qlo = \frac{\sN 0}{2\,(\vi(q))^N}
   - \frac{(\vi(q))^N}{2\,(\vi(q^2))^N} \, \labl{cnsJ}
and
  \be  \chim m(q) \equiv \trNS \Ppm m\,\qlo
   \!\!  = \displaystyle \intt  \eE^{\mp\ii mt} 
   \LLb \frac{\wi(q;-\eit q^{1/2})}{\vi(q)} \LRb^N_{}\, 
   = \frac{\sN m}{(\vi(q))^N} \labl{chimNS}
for $m\iN\natnum$.
(Note that the latter result does not depend on whether $\Pp m$ or $\Pm m$
is used, since $\sN m = \sN {-m}$.)

Expressing the integer $m$ either as $ m=nN+j$ or as $m=(n+1)N-j$ with
$1\le j\le\el$, by shifting the summation indices we obtain the relation
  \Be \sN {nN+j} = q^{nj+n^2N/2}_{}\, \sN j . \Ee
Hence we have
  \be  \chim{nN+j}(q) = q^{nj+n^2N/2}_{}\, \chim j(q) \,; \labl{n+}
in the same manner we obtain
  \Be  \chim{(n+1)N-j}(q) = q^{n(N-j)+n^2N/2}_{}\, \chim{N-j}(q) , \Ee
or alternatively
  \be  \chim{(n+1)N-j}(q) = q^{-(n+1)j+(n+1)^2N/2}_{}\, \chim{j}(q) \,.\labl{n-}
For $j=0$ we have instead
  \be  \chim{nN}(q) = q^{n^2N/2}_{}\, [\chio(q)+\chiJ(q)]  \ee
for all $n>0$.
  
\subsection{\sonhz\ characters for even $N$}

When we use the information about the \hwv s \wrt the affine \lie\ \sonh\ at 
level 2 that we obtained above,
we can derive the characters of the \ihwm s of \sonhz\ 
by comparing the decomposition \erf{Hdec} with the decompositions \erf{anse}
and \erf{anso}. We first consider the case $N=2\el$. 

By comparison of \erf{Hdec} with \erf{anse} we find 
  \be \bearll
  \chicjl (q)\, \chizj (q) \!\!&= \chim j (q) + \chim {N-j} (q)
    + \chim {N+j} (q) + \chim {2N-j} (q) + \ldots \\[.3em]
  &\equiv \dstyle\sumni 0 [\chim {nN+j} (q) + \chim {(n+1)N-j} (q)] 
  \eear \ee
for even $j$. Using \erf{n+} and \erf{n-}, this becomes
  \be \bearll  \chicjl (q)\, \chizj (q) \!\!&
   = \chim j (q) \dstyle\sumnZ q^{nj+n^2N/2} \\[.3em]
  &= q^{-j^2/2N}\, \psinh j (q)\,\chim j (q) 
   = q^{-j^2/2N}\, \psinh j (q)\, \Frac{\sN j}{(\vi(q))^N} 
  \,. \eear \ee
Analogously, with \erf{anse} we obtain the same result for odd $j$.
By inserting the coset characters $\chicjl$ \erf{chiCj} we then get
  \be \chizj (q) = q^{-j^2/2N}\, \Frac{\sN j}{(\vi(q))^{N-1}}
  \,.  \labl{chizj}
For $j=\el$ one has to read this result with \erf{chizs}, which
means that our result only describes the sum of the irreducible
characters $\chizs$ and $\chizc$. By comparison with \erf{chimNS},
we may also rewrite the result in the form 
  \be \chim {nN+j} (q) = \frac{q^{(nN+j)^2/2N}}{\vi(q)} \, 
  \chizj (q)\,,   \qquad \chim {(n+1)N-j} (q) = 
  \frac{q^{((n+1)N-j)^2/2N}}{\vi(q)} \, \chizj (q) \labl{zvir}
for $j=\onetol$.

Comparing \erf{Hdec} again with \erf{anse}, we also find
  \be \hsp{-.7} \bearll
  \chicol(q)\, \chizo(q) + \chicvl(q)\, \chizv(q) \!\!\!&=
  \chio(q)  + \dstyle\sumni 1 \chim {nN} (q) \\[.4em]
  &=[\chio(q) + \chiJ(q)]\, \LLb \half 
    + \half\psinh0 (q) \LRb - \chiJ(q) \\[.7em]
  &= \half[\chio(q)-\chiJ(q)]+\half\psinh0 (q) \, [\chio(q)+\chiJ(q)] \\[.6em]
  &= \Frac{(\vi(q))^N}{2\,(\vi(q^2))^N} +
    \psinh0(q) \, \Frac{\sN 0}{2\,(\vi(q))^N} \eear \labl{coo}
and
  \be \bearll \chicol(q)\, \chizv(q) + \chicvl(q)\, \chizo (q) \!\!\!
  &= \chiJ (q)  + \dstyle\sumni 1 \chim {nN} (q) \nline8
  &= - \Frac{(\vi(q))^N}{2\,(\vi(q^2))^N} +
    \psinh0(q) \, \Frac{\sN 0}{2\,(\vi(q))^N}\,. \eear \hsp4 \labl{cov}
Subtraction of \erf{cov} from \erf{coo} yields 
  \be  
  [\chicol(q) - \chicvl(q)]\cdot[\chizo(q) - \chizv(q)]   
  = \LLb \Frac{\vi(q)}{\vi(q^2)} \LRb^N  \equiv \chio (q) - \chiJ (q) 
  \,,  \ee
so that by inserting \erf{chiCovm} we obtain 
  \be  \chizo(q) - \chizv(q) = 
  \LLb \Frac{\vi (q)}{\vi (q^2)} \LRb^{N-1}  \,. \labl{dif}
Analogously, by adding \erf{coo} and \erf{cov} we get
  \be  
  [\chicol(q) + \chicvl(q)]\cdot[\chizo(q) + \chizv(q)]  
  = \psinh0(q)\, \frac{\sN 0}{(\vi(q))^N} \,, 
  \ee
and hence inserting \erf{chiCovp} we obtain
  \be \chizo (q) + \chizv (q) = \frac{\sN 0}{(\vi(q))^{N-1}} \,. \labl{sum}
In summary, we have derived that
  \be  \bearl
  \chizo(q) = \half \Llb \Frac{\sN 0}{(\vi(q))^{N-1}} + 
     \LLb \Frac{\vi (q)}{\vi (q^2)} \LRb^{N-1} \Lrb
  \equiv \Frac1{2\,(\vi(q))^{N-1}} \,\llb \sN 0 + (\psiemq)^{N-1} \lrb\,,
  \\{}\\[-.5em]
  \chizv(q) = \half \Llb \Frac{\sN 0}{(\vi(q))^{N-1}} - 
     \LLb \Frac{\vi (q)}{\vi (q^2)} \LRb^{N-1} \Lrb
  \equiv \Frac1{2\,(\vi(q))^{N-1}}\,\llb \sN 0 - (\psiemq)^{N-1} \lrb\,.
  \eear \labl{both}

Further, comparison with \erf{cnso} and \erf{cnsJ} yields
  \be \chio (q) + \chiJ (q) = \frac1{\vi(q)} \,
  [ \chizo (q) + \chizv (q) ] \,, \labl{cOJ}
while comparison with \erf{chimNS} and \erf{n+} shows that
  \be \chim {nN} (q)  = \frac{q^{n^2N/2}}{\vi(q)} \,
  [ \chizo (q) + \chizv (q) ] \,. \labl{cNSnN} 

\subsection{\sonhz\ characters for odd $N$}

Now we consider the case $N=2\el+1$. {}From \erf{Hdec} and \erf{anso} we find 
  \be  \hsp{-1.3} \bearll
  \chicjzN {2j} (q)\, \chizj(q) \!\!\!&= \chim j (q) + \chim {2N-j} (q)
    + \chim{2N+j}(q) + \chim{4N-j}(q) + \chim{4N+j}(q) + \ldots \nline7
  &\equiv \dstyle\sumni 0 [\chim {2nN+j} (q) + \chim {2(n+1)N-j} (q)] \nline7
  &= \chim{j}(q) \dstyle\sumnZ q^{2nj+2n^2N}  
   = q^{-j^2/2N}\,\psinz {2j}(q)\, \chim{j}(q) \eear \ee
for $j$ even, and 
  \be  \hsp{-.3} \bearll
  \chicjzN {2N-2j} (q)\, \chizj(q) \!\!\!
  &= \dstyle\sumni 0 [\chim{(2n+1)N+j}(q) + \chim{(2n+1)N-j}(q)] 
   = \chim j (q) \dstyle\sumnZ q^{-(2n+1)j+(2n+1)^2N/2} \nline7
  &= q^{-j+N/2} \chim j (q) \, \dstyle\sumnZ q^{2n(N-j)+2n^2N} 
   = q^{-j^2/2N} \, \psinz {2N-2j} (q) \, \chim j (q) \eear \ee
for $j$ odd.
By inserting the coset characters \erf{chiCj} we then arrive once again at
the formul\ae\ \erf{chizj} and \erf{zvir} for $j=\onetol$. 

In the same manner we find
  \be  \hsp{-1.2} \bearll 
  \chicozN(q)\, \chizo(q) + \chicvzN(q)\, \chizv(q) \!\!\!&
   = \chio(q)  + \dstyle\sumni 1 \chim {2nN} (q) \nline7
  &= [\chio(q)+\chiJ(q)]\, \LLb \half
  +\half \dstyle\sumnZ q^{2n^2N} \LRb -\chiJ(q) \nline7
  &= \Frac{(\vi(q))^N}{2\,(\vi(q^2))^N} +
  \psinz 0(q) \, \Frac{\sN 0}{2\,(\vi(q))^N} \eear \ee 
and
  \be  \bearll  \chicozN(q)\, \chizv(q) + \chicvzN(q)\, \chizo(q) \!\!&
   = \chiJ (q)  + \dstyle\sumni 1 \chim {2nN} (q) \nline7 
  &= - \Frac{(\vi(q))^N}{2\,(\vi(q^2))^N} +
  \psinz 0(q) \, \Frac{\sN 0}{2\,(\vi(q))^N} \,. \eear \hsp{3.5} \ee 
Thus we also obtain again the relations \erf{dif} and \erf{sum} for
$\chizo$ and $\chizv$, and hence also \erf{both} and \erf{cNSnN}.

\Sect{Summary and outlook}c

\subsection{Decomposition of the tensor product}

Let us now summarize some of our results on the tensor product decompositions. 
To this end we first note that $q^\dh/\vi(q)$ is precisely the character 
of the Verma module $M(c,\dh)$ of the Virasoro \alg. For central charge
$c=1$ the Verma module $M(c,\dh)$ is irreducible as
long as $4\dh\neq m^2$ for $m\iN\zet$; otherwise there exist null
states. The characters of the irreducible modules
$V(1,\dh)$ of the $c=1$ Virasoro algebra are then given by
  \be  \chivir\dh(q) = \left\{ \bearll (\vi(q))^{-1}\,
  [q^{m^2/4}-q^{(m+2)^2/4}] \quad & \mbox{if}\
  \dh=\frac{m^2}4\ {\rm with}\ m\iN\zet\,, \\{}\\[-.8em]
  (\vi(q))^{-1}\,q^\dh & \mbox{otherwise.} \eear \right. \ee
Thus for $4\dh=m^2$ with $m\iN\zet$ the Verma module
character can be decomposed as follows:
  \be \frac{q^{m^2/4}}{\vi(q)} = \frac1{\vi(q)} 
  \sum_{k=0}^\infty \,\LLb q^{(m+2k)^2/4} - q^{(m+2k+2)^2/4} 
  \LRb = \sum_{k=0}^\infty \chivir {(m+2k)^2/4}(q) \,. \labl{devir}
Correspondingly we write 
  \be  \Reh \dh := \left\{ \bearll \dstyle\bigoplus_{k=0}^\infty 
  V(1,\Frac{(m+2k)^2}4) \qquad & \mbox{if}\ \dh=\Frac{m^2}4\
  \mbox{with}\ m\in\zet\,, \nline6
  V(1,\dh) & \mbox{otherwise.} \eear \right. \labl{deR}

Using also the formul\ae\ \erf{ccwa} and \erf{ccwb} for the coset conformal 
weights, we can summarize our results of \srf9
by the following description of the big Fock space $\HNSh$. 
Recalling the decomposition 
  \be \HNSh = \Hho \oplus \Hhj \oplus
  \bigoplus_{m=1}^\infty (\Hhm \otimes \complex^2)  \ee
of $\HNSh$ into $\fAg$-sectors, we can express the splitting of $\HNSh$
into tensor products of the Virasoro modules \erf{deR}
and the irreducible highest weight modules of \sonhz\ (that is, $\Uo,\,\Uv,\,
\Uj$, and also $\Us$ and $\Uc$ when $N=2\el$) as follows. Our results show that
  \be \Hhh {nN} = \LLb \Uo\oplus\Uv \LRb \,\otimes\,  
  \Reh{\Delc\circ} \,, \labl{decHN}
for $n=1,2,...\,$, as well as
  \be \begin{array}{r}
  \Hhh {nN+j} = \Uj \,\otimes\, \Reh{\Delc j}\,, \nline4 
  \Hhh {(n+1)N-j} = \Uj \,\otimes\, \Reh{\Delcb j}\,\, \eear \labl{decHj}
for $n=0,1,...\,$ and $j=\onetolme$. When $N=2\el+1$, \erf{decHj} also holds 
for $j=\el$, while for $j=\el$ and $N=2\el$ we have
  \be \Hhh {nN+\el} = \LLb \Us \oplus \Uc \LRb  \,\otimes\, 
  \Reh{\Delc\rms}  \labl{decHsc}
for $n=0,1,...\,$. Note that the modules $\Reh\dh$ appearing in the 
decompositions \erf{decHN}, \erf{decHj} and \erf{decHsc} are all irreducible 
as long as $\sqrt{2N}\notin\natnum$. Otherwise we can write
$N=2K^2$ with $K\iN\natnum$, and then the modules
$\Reh{\Delc\circ}$ and $\Reh{\Delc j}$, $\Reh{\Delcb j}$
with $j=mK$, $m=1,2,...\,$ and $j\le\el$, split up as in \erf{deR}.

Besides the coset Virasoro generators,
the chiral symmetry algebra of the orbifold coset theory contains further
operators \cite{dvvv}. The observation above implies in particular that 
when acting on $\fAg$-sectors other than $\Hho$ and $\Hhj$,
for $\sqrt{2N}\notin\natnum$ all these additional generators make transitions 
between the sectors of the gauge invariant fermion \alg\ $\fAg$;
for $N=2K^2$ ($K\iN\natnum$) the additional generators generically still 
make transitions, except that they can map sectors with $j=mK$ to themselves.
It follows in particular that we can distinguish between elements of the
coset Virasoro \alg\ and elements of the full coset chiral \alg\ which are
not contained in the coset Virasoro \alg\ by acting with them on suitable
$\fAg$-sectors.

\subsection{The sectors $\Hho$ and $\Hhj$}

It still remains to analyze the decomposition of the $\fAg$-sectors $\Hho$ and
$\Hhj$ explicitly. From \erf{cOJ} we conclude that
  \be \Hho \oplus \Hhj = \LLb \Uo\oplus\Uv \LRb \,\otimes\,  \Reh 0 \,. \ee
Now $\Reh 0$ is always reducible, independent of the particular value
of the integer $N$. We claim that
  \be \bearl
  \Hho = \Hhzo \otimes \dstyle\bigoplus_{k=0}^\infty V(1,(2k)^2)
  \,\,\,\oplus\,\,\, \Hhzv \otimes 
  \dstyle\bigoplus_{k=0}^\infty V(1,(2k+1)^2) \,,\\{}\\[-.6em]
  \Hhj = \Hhzo \otimes \dstyle\bigoplus_{k=0}^\infty V(1,(2k+1)^2)
  \,\,\,\oplus\,\,\, \Hhzv \otimes 
  \dstyle\bigoplus_{k=0}^\infty V(1,(2k)^2) \,. \eear \ee
This can be seen by decomposing the characters $\chio$ and $\chiJ$ as follows:
  \be  \bearll
  \chio (q) \!\!& =  \Frac{\sN 0}{2(\vi(q))^N} +
  \Frac{(\vi(q))^{N-2}}{2(\vi(q^2))^{N-1}}
  \dstyle\sum_{k\in\zet} (-1)^k q^{k^2} \nline7 
  & \equiv \Frac{\sN 0}{2(\vi(q))^N} +
  \Frac{(\vi(q))^{N-2}}{2(\vi(q^2))^{N-1}}
  \dstyle\sum_{k=0}^\infty \llb q^{(2k)^2} - 2 q^{(2k+1)^2}
  + q^{(2k+2)^2} \lrb \nline7 
  & = \chizo (q) \Frac1{\vi(q)} \dstyle\sum_{k=0}^\infty 
  \llb q^{(2k)^2} -  q^{(2k+1)^2} \lrb +
  \chizv (q) \Frac1{\vi(q)} \dstyle\sum_{k=0}^\infty 
  \llb q^{(2k+1)^2} -  q^{(2k+2)^2} \lrb \nline7 
  & \equiv \chizo (q) \cdot \dstyle\sum_{k=0}^\infty \chivir {(2k)^2} (q)   + 
  \chizv (q) \cdot \dstyle\sum_{k=0}^\infty \chivir {(2k+1)^2} (q)
  \,. \eear \ee
(In the first line we used \erf{76}.) Similarly,
  \be  \chiJ = \chizo \cdot \sum_{k=0}^\infty \chivir {(2k+1)^2}
   + \chizv \cdot \sum_{k=0}^\infty \chivir {(2k)^2} \,. \hsp4 \ee
It follows that besides $\OmfOo 0 \equiv\OM$ and $\OmfJv 0\equiv\Ovvo$, 
there must exist further simultaneous \hwv s of $\sonhz$ and the coset 
Virasoro algebra, namely, for $k=0,1,2,...\,$, \hwv s
$\OmfOo {2k+2},\,\OmfOv {2k+1} \iN\Hho$ and 
$\OmfJo {2k+1},\,\OmfJv {2k+2} \iN\Hhj$,
with \sonhz-weights $\lo$, $\lv$, $\lo$, $\lv$, \resp, and with coset
conformal weights $(2k+2)^2$, $(2k+1)^2$, $(2k+1)^2$, $(2k+2)^2$, \resp. 
Those vectors with unit coset conformal weight have a relatively simple form;
we find
  \be  \hsp{-.7} \OmfJo 1 = \left\{ \bearll \! \dstyle\sum_{k=1}^\el
  \lLb \xb k\mh \y k\mh - \yb k\mh \x k\mh \lRb \, \OM & \forzl\,, \nline7
  \!\! \Llb \dstyle\sum_{k=1}^\el\!\lLb \xb k\mh \y k\mh - \yb k\mh \x k\mh \lRb
  + \xb {\el+1}\mh \yb {\el+1}\mh \Lrb \, \OM & \forzle\,, \eear \right. \ee
as well as
  \be   \OmfOv 1 = \x 1\mh \y 1\mh \, \OmfJo 1 +
  (\x 1{-3/2} \y 1\mh + \y 1{-3/2} \x 1\mh) \, \OM \,.  \hsp4 \ee
In contrast, the \hwv s with larger coset conformal weight are more difficult
to identify.

\subsection{\sonhz\ characters}

Our idea to employ the \rep\ theory of the gauge group \Oz\ allowed us to 
deduce simple formul\ae\ for the characters of the (\NS sector) \ihwm s of 
\sonh\ at level two. They are given by the 
expressions \erf{chizj} for $\chizj$ and \erf{both} for $\chizo$ and $\chizv$.
Note that, not surprisingly, these results have a simple functional
dependence on the integer $N$, even though the details of 
their derivation (involving e.g.\ the relation with the orbifold
coset theory) depend quite non-trivially on whether $N$ is even or odd.

Our results for these characters are not new. In \cite{scya5}, the conformal 
embedding of \sonhz\ into \unh\ at level one was employed to identify
(sums of) \sonhz\ characters with characters of \sunhe. Indeed, the
restricted summation over the lattice vector $\vecm\iN\zet^N$ in the
formula \erf S for $\sN m$ precisely corresponds to the summation over
the appropriately shifted root lattice of \sun. 

With the help of the
conformal embedding only the linear combination $\chizo+\chizv$ of the
irreducible characters $\chizo$ and $\chizv$ is obtained, which is just
the level-one vacuum character of \sunh. However, the orthogonal linear 
combination $\chizo-\chizv$ is known as well; it has been obtained in 
\cite[p.\,233]{kawa} by making use of the theory of modular forms.

\subsection{A homomorphism of fusion rings}

In \srf8 we were able to identify the \sonhz\ \hwm s within the sectors of 
the intermediate \alg\ $\fAg$ which are governed by the gauge group \Oz.
Our results amount to the following assignment $\rho$ of
the \Oz-representations $\Phi$ to the WZW sectors $\phi$\,:
  \be \bearl
  \rho (\rpo) = \pfo \,,\qquad \rho (\rpj) = \pfv \,,\\{}\\[-.4em]
  \rho (\rp {(n+1)N}) = \pfo + \pfv \,,\\{}\\[-.4em]
  \rho (\rp {nN+j}) = \rho (\rp {(n+1)N-j}) = \pfj j 
  \qquad {\rm for}\ j=\onetolme \,,\\{}\\[-.6em]
  \rho (\rp {nN+\el}) = \rho (\rp {(n+1)N-\el}) =
  \left\{ \begin{array}{cl} \pfs + \pfc  & \forzl\,, \\[.3em]
  \pfj \el & \forzle\,, \eear \right. \eear \ee
for $n=0,1,2,...$
(Note that in the case of $\rpo$ and $\rpj$, the action of $\rho$ does not
directly correspond to the decomposition of the $\fAg$-sectors into \sonhz\ 
sectors.)

The multiplication rules of the \rep\ ring \Ro\ of \Oz\ are given by the
relations \erf{tp}. The level-two WZW sectors generate a fusion ring, too,
which we denote by \Rw. The ring \Rw\ has a fusion subring \Rn\ which is 
generated
by those primary fields which appear in the \NS sector $\HNSh$. The fusion 
rules, i.e.\ the structure constants of \Rw, can be computed with the help of 
the Kac\hy Walton and Verlinde formul\ae\ (see e.g.\ \cite{fuva3}).
For the subring \Rn\ one finds the tensor product decompositions listed
in appendix \ref0.

Inspection shows that \Rn\ is in fact isomorphic to the \rep\ ring of the 
dihedral group \Dn. Now for any $N$ the group \Dn\ is a finite subgroup of
\Oz. As a consequence, the mapping $\rho$ actually constitutes a fusion ring
{\em homomorphism\/} from the \rep\ ring \Ro\ of \Oz\ to the fusion 
subring \Rn\ of \Rw. (It is also easily checked that for odd $N$ the
homomorphism $\rho$ is surjective, while for even $N$ the image does not contain
the linear combination $\pfs-\pfc$.) This observation explains to a certain 
extent why, in spite of the fact that the WZW observable algebra \AW\ is much
smaller than the \Oz-invariant \alg\ $\fAg$, the group \Oz\ nevertheless
provides a substitute for the gauge group in the DHR sense. But even in view of
this relationship it is still surprising how closely the WZW 
superselection structure follows the representation theory of \Oz.

One may speculate that the presence of the homomorphism $\rho$ indicates 
that the gauge group \Oz\ is in fact part of the full (as yet unknown)
quantum symmetry of the \wzwt\ that fully takes over the \role\ of the DHR
gauge group. This is possible because all sectors in the \NS part of the
\wzwt\ have integral quantum dimension. Now in rational \cft\
sectors with integral quantum dimension are actually extremely
rare. It will be interesting to study the relationship between the \rep\ ring
of \Ok\ or \Uk\ and the WZW fusion ring in more general cases where (most of)
the WZW sectors possess non-integral quantum dimension.


\newpage
\appendix\sect{Appendix}

\subsection{Commutators of the fermion modes $\xpm kr$ with the currents}

By direct calculation, we obtain
  \be  [\HH j,\xpm kr]=\del jk\,\xpm kr \,, \qquad [\HH j,\xbpm kr]=
  -\del jk\,\xbpm kr \,,   \labl{Hx}
for all $j,k=\onetol$, and similarly, $\forzle$,
  \be  [\HH j , \xbpm {\el+1}r] = 0  \labl{Hxl}
for all $j=\onetol$.

To find also the commutators of the fermion modes with the raising operators
$\EE j+$, we first compute
  \be  [\Jmtee ij,\cpm kr] = \half\eps\,(\eta\mp1)\,\del jk\,\ceps i{m+r}
  - \half\eta\,(\eps\mp1)\,\del ik\,\ceta j{m+r}  \,. \labl{Jc}
Analogous relations hold for $[\Jmtee ij,\cbpm kr]$. 
When $N=2\el+1$ we have in addition the relation
  \Be [ \Jmtee ij , \b {2\el+1}qr ] = 0  \Ee
and 
  \be \bearl
  [\Jmttp j,\cpm kr] = \mp \,\del jk \, \b{2\el+1}1{m+r} \,, \qquad \
  [\Jmttm j,\cpm kr] = 0 \,, \\[.5em]
  [\Jmttpm j,\b {2\el+1}1r] = - \, \cpm j{m+r} \,,  \eear  \labl{Jc3}
and similar relations for $[\Jmttp j,\cbpm kr]$,
$[\Jmttm j,\cbpm kr]$ and $[\Jmttpm j,\b{2\el+1}2r]$.
{}From these results we learn that
  \be  \bearl
  [\EE 0+,\xpm kr] = \del k2\, \xbpm 1{r+1} - \del k1\, \xbpm 2{r+1}\,, \qquad\
  [ \EE 0+ , \xbpm kr ] = 0\,, \\{}\\[-.5em]
  [\EE j+,\xpm kr] = -\del k{j+1}\, \xpm jr\,, \qquad \
  [\EE j+,\xbpm kr]= \del kj\, \xbpm {j+1}r \quad{\rm for}\ j=\onetolme \,, 
  \\{}\\[-.6em]
  [\EE \el+,\xpm kr] = 0\,, \qquad [\EE \el+,\xbpm kr] = \left\{ \bearll
  \del k\el\, \xpm {\el-1}r-\del k{\el-1}\, \xpm\el r & \forzl\,, \nline8
  \del k\el\, \xbpm{\el+1}r-\del k{\el+1}\, \xpm\el r & \forzle \,. 
  \eear\right.  \eear \labl{eox}

Taking into account that $(\xpm jr)^2 = (\xbpm jr)^2 = 0$ and $(\xpm jr)^* = 
\xbmp j{-r}$, these relations imply that
  \be  \bearl [\EE j+,\Xpm kr] = 0 \qquad {\rm for}\,\, j=\onetol \,, \\[.5em]
  [ \EE j+,\Xbpm kr ] = 0  \qquad {\rm for}\,\, j=\onetolme\,.\eear \labl{EX}
For $j=\el$ we have instead
  \be \bearl
  [\EE\el+,\Xbpm kr]\cdot\Xpm\el r = 0 \qquad \forzl\,, \\[.7em]
  [\EE\el+,\Xbpm kr]\cdot\xbpm{\el+1}r = 0 \quad {\rm and} \quad
  [\EE\el+,\xbpm{\el+1}r]\cdot \Xpm\el r = 0 \quad {\rm for}\ N=2\el+1 \,. 
  \eear \labl{ElXb}
Finally, for $j=0$ we find
  \be  [\EE0+,\Xbpm kr] = 0 \,, \qquad
  [\EE0+,\Xpm kr] \cdot \Xbpm0{r+1} = 0  \,. \labl{E0X}

\subsection{The action of the gauge group \Oz}

For the Fourier modes $\cpm jr$ and $\cbpm jr$ 
the actions \erf{sgamt} of $\sgamt$, $t\iN\reals$,
and \erf{seta} of $\seta$ read
  \be  \bearll
  \sgamt(\cpm jr)=\cost \cpm jr - \sint \cbpm jr \,,\ &
  \sgamt(\cbpm jr)=\sint \cpm jr + \cost \cbpm jr \,,\nline5
  \seta(\cpm jr)=\cpm jr\,, & \seta(\cbpm jr)=-\cbpm jr  \,, \eear \labl{Oz}
so that the combinations $\xpm jr$ transform as
  \be  \sgamt(\xpm jr) = \eE^{\pm \ii t}\,\xpm jr \,,\quad
    \seta(\xpm jr) =\xmp jr \,.  \hsp1 \ee
Analogously,
  \be  \sgamt(\xbpm jr) = \eE^{\pm\ii t}\,\xbpm jr \,,\quad
  \seta(\xbpm jr) =\xbmp jr\,. \hsp1 \ee
Hence the combinations $\Xpm jr$ transform as
  \be  \sgamt(\Xpm jr) = \eE^{\pm\ii jt}\,\Xpm jr \,,\quad
  \seta(\Xpm jr) =\Xmp jr\,, \hsp1 \ee
and analogously,
  \be  \sgamt(\Xbpm jr) = \eE^{\pm\ii(\el-j)t}\,\Xbpm jr \,,\quad
  \seta(\Xbpm jr) =\Xbmp jr  \,. \hsp1 \labl{OZ}

\subsection{The fusion rules of \sonhz}\label0

In this appendix we present the relations of the fusion ring $\Rn\subset\Rw$,
i.e.\ the fusion rules for those primary fields of 
the \wzwt\ based on \sonhz\ which correspond to the \sonhz\ highest weight 
modules that appear in the \NS sector. For $N=2\el$ we have
  \be  \bearl
  \pfv \fstar \pfv = \pfo\,, \qquad\quad \pfv \fstar \pfs = \pfc \,, \nline7
  \pfs \fstar \pfs = \pfc \fstar \pfc = \left\{ \bearll
       \pfo & {\rm for}\ \el\iN2\zet\,, \\[.2em] 
       \pfv & {\rm for}\ \el\iN2\zet+1\,, \eear\right. \nline7
  \pfs \fstar \pfc = \left\{ \bearll
       \pfv & {\rm for}\ \el\iN2\zet\,, \\[.2em] 
       \pfo & {\rm for}\ \el\iN2\zet+1\,, \eear\right. \nline6
  \pfv \fstar \pfj j = \pfj j\,,  \qquad\quad
  \pfs \fstar \pfj j = \pfc \fstar \pfj j = \pfj{\el-j}\,, \nline4
  \pfj i \fstar \pfj j = \pfj{\,|i-j|\,} + \pfj{i+j} \,.
  \eear \ee
Here it is to be understood that whenever on the right hand side a label
$j$ appears which is larger than $\el$, it must be interpreted as the number 
  \be  j' := N-j \,, \ee 
and when the label equals zero or $\el$, one has to identify $\pfj j$ as
the sum
  \be  \pfj0 \equiv \pfo + \pfv\,,  \qquad  \pfj\el \equiv \pfs + \pfc \,. \ee

For $N=2\el+1$ the fusion rules read 
  \be  \bearl
  \pfv \fstar \pfv = \pfo\,,  \qquad\quad
  \pfv \fstar \pfj j = \pfj j \,, \nline6
  \pfj i \fstar \pfj j = \pfj{\,|i-j|\,} + \pfj{i+j} \,.
  \hsp4  \eear \ee
This time it is understood that when $j$ is larger than $\el$,
it stands for the number $j' := N-j$, and again that $\pfj0 \equiv \pfo + \pfv$.

The \NS sector fusion rules which are not listed explicitly all follow from 
the commutativity and the associativity of the fusion product and from the 
fact that $\pfo$ is the unit of the fusion ring.

\bigskip\medskip\small
\noindent{\bf Acknowledgement.}\\ We thank K.-H.\ Rehren for a careful
reading of the manuscript, and K.\ Fredenhagen and K.-H.\ Rehren for helpful 
discussions.

  \newcommand\wb       {\,\linebreak[0]} \def\wB {$\,$\wb}
  \newcommand\Bi[1]    {\bibitem{#1}}
  \newcommand\Erra[3]  {\,[{\em ibid.}\ {#1} ({#2}) {#3}, {\em Erratum}]}
  \newcommand\BOOK[4]  {{\em #1\/} ({#2}, {#3} {#4})}
  \renewcommand\J[5]   {\ {\sl #5}, {#1} {#2} ({#3}) {#4} }
  \newcommand\JJ[5]    {{\sl #5}  {#1} {#2} ({#3}) {#4}}
  \newcommand\Prep[2]  {{\sl #2}, preprint {#1}}
  \newcommand\Prew[2]  {{\sl #2}, {#1} preprints}
 \def\adma  {Adv.\wb Math.}
 \def\jf    {J.\ Fuchs}
 \def\coma  {Con\-temp.\wb Math.}
 \def\ijmp  {Int.\wb J.\wb Mod.\wb Phys.\ A}
 \def\lemp  {Lett.\wb Math.\wb Phys.}
 \newcommand\npbF[5] {{\sl #5}, \nupb\ {#1} [FS{#2}] ({#3}) {#4}}
 \def\npbp  {Nucl.\wb Phys.\ B (Proc.\wb Suppl.)}
 \def\nupb  {Nucl.\wb Phys.\ B}
 \def\comp  {Com\-mun.\wb Math.\wb Phys.}

 \def\A       {Algebra}
 \def\alg     {algebra}
 \def\Be     {{Berlin}}
 \def\BIR    {{Birk\-h\"au\-ser}}
 \def\Ca     {{Cambridge}}
 \def\Con     {Conformal\ }
 \def\cua     {current algebra}
 \def\CUP    {{Cambridge University Press}}
 \def\fts     {field theories}
 \def\furu    {fusion rule}
 \def\GB     {{Gordon and Breach}}
 \newcommand{\inBO}[7]  {in:\ {\em #1}, {#2}\ ({#3}, {#4} {#5}), p.\ {#6}}
 \def\Infdim  {Infinite-dimensional}
 \def\KLU    {{Kluwer Academic Publishers}}
 \def\NY     {{New York}}
 \def\oa      {operator algebra}
 \def\q       {quantum\ }
 \def\Q       {Quantum\ }
 \def\con     {conformal\ }
 \def\qg      {quantum group}
 \def\modinv  {modular invarian}
 \def\inv     {invariance}
 \def\Rep     {Representation}
 \def\SV     {{Sprin\-ger Verlag}}
 \def\sym     {symmetry}
 \def\syms    {sym\-me\-tries}
 \def\trfo    {transformation}
 \def\wzw     {WZW\ }
 \def\WZW     {Wess\hy Zu\-mino\hy Wit\-ten\ }

\end{document}